\documentclass[preprint,12pt]{elsarticle}
\usepackage[margin=1.8cm]{geometry}
\usepackage{amsmath}
\usepackage{amsfonts}
\usepackage{amssymb}
\usepackage{graphicx}
\usepackage{lmodern}
\usepackage{hyperref}
\usepackage{dcolumn}
\usepackage{bm}
\usepackage[mathlines]{lineno}
\usepackage{float}
\usepackage[T1]{fontenc}
\usepackage{subfigure}

\journal{Physics Letters A}

\usepackage{numcompress}

\begin{document}

\begin{frontmatter}

\title{\textbf{Confined Neutrino Oscillation}}



\author[mymainaddress]{Pralay Chakraborty}
\ead{pralay@gauhati.ac.in}

\author[mymainaddress]{Subhankar Roy\corref{mycorrespondingauthor}}
\cortext[mycorrespondingauthor]{Corresponding author}
\ead{subhankar@gauhati.ac.in}

\address[mymainaddress]{Department of Physics, Gauhati University, India}

\begin{abstract}

A thought experiment is designed to speculate on the neutrino being trapped in an impenetrable potential well. Considering both relativistic and non-relativistic scenarios, we delve into several interesting facets connected to the flavour oscillation.

\end{abstract}

\begin{keyword}
Neutrino Oscillation \sep Infinite Square Well Potential\sep Non-Relativistic Case\sep Relativistic Case.
\end{keyword}

\end{frontmatter}

\section{Introduction \label{Introduction}}

The idea of neutrino was first introduced by Wolfgang Pauli in the year of 1930. Initially it was assumed that hardly it could be detected in the experiments, but later in the year of 1956 it was detected experimentally. The neutrinos belong to the lepton family and they exhibit peculiar characteristics which cannot be understood with the help of present understanding of Standard Model\,(SM)\,\cite{Glashow:1961tr, Weinberg:1967tq, Herrero:1998eq} of Particle physics. For example, why the neutrinos posses mass is still unanswered in the framework of SM. But from the quantum mechanics point of view, if the neutrino is  massive, then it must exhibit a peculiar phenomenon called neutrino oscillation \cite{Farzan:2017xzy, Ellis:2020hus, Fantini:2018itu} which says that the neutrino with certain flavour changes its identity from one to another while it travels in space, and this prediction agrees well to the experimental observation \cite{Super-Kamiokande:1998qwk, Super-Kamiokande:1998kpq, SNO:1999crp, SNO:2002tuh, SNO:2002hgz}. So far, we know that the neutrinos participate only in weak interaction and thus the Universe is almost transparent to the neutrinos. So, as per the present understanding of neutrinos and their interaction is concerned, we understand that hardly, there exists a possibility to trap a neutrino. However, extremely compact stars such as neutron stars are characterized by the presence of trapped null geodesics, which are specific paths followed by subatomic particles like photons and neutrinos, in the curved space-time. When the neutron stars enter a phase in their evolutionary cycle that permits geodetical motion of neutrinos, it may so happen that inside the neutron star a certain portion of the neutrinos produced within their dense interior becomes trapped or confined within the star \cite{Stuchlik:2011zzb}. These trapped neutrinos are unable to escape the strong gravitational field of these dense objects. In such extreme conditions, the interactions between neutrinos become highly influential \cite{Mirizzi:2015eza}. These interactions can trigger flavour oscillations, leading to a collective behaviour where neutrinos of different energies and flavours synchronize and oscillate together as a group \cite{Duan:2006jv, Duan:2006an}. In the present work, we set a thought experiment to explore the consequences on neutrino oscillation if somehow the neutrino could have been confined within a certain potential well. The findings from this study though speculative and based on hypothetical circumstances might be relevant in the context of above mentioned systems.


The theoretical formulation of the probability of conversion from one flavour to another for neutrino was first proposed by Bruno Pontecorvo in the year of 1957 \cite{Pontecorvo:1957cp}, based on the principles of quantum mechanics. However, the theory holds good for high energetic relativistic free neutrinos \cite{Bilenky:1976yj, Bilenky:1978nj}. In the present work, as we assume that the neutrino is bound within certain potential well and hence we shall explore the possibilities concerning both: non-relativistic and relativistic scenarios. However, for the latter case, the situation is little difficult to deal with because of the appearance of the Klein Paradox \cite{Calogeracos:1998rf, Dombey:1999id, Hansen:1980nc, Su:1993kn, Calogeracos:1999yp} in the bound system. What we expect is that, being bound, the neutrino is available to stay in any of the energy states allowed by the system. As a result, it is expected that the Probability of conversion from one flavour to another will certainly be affected by the quantised energies of the neutrino mass eigenstates.

The plan of the paper is given as follows: In section \ref{section 2}, we discuss the flavour mixing and the parametrization of the PMNS matrix. We study the neutrino oscillation probability inside an infinite square well potential in section \ref{section 3}. The section \ref{section 4} is devoted to some observations regarding the present work. We write the summary and discussion of our work in section \ref{section 5}.

\section{Parametrization of PMNS matrix \label{section 2}}

The quantum mechanics says that the neutrino mass eigenstates are not equivalent to the neutrino flavour states, rather the mass eigenstates mix to produce the neutrino flavour states. Experimentally, the flavour states are observable and not the mass eigenstates. The neutrino mass eigenstates, $\nu_{i=1,2,3}$ carry definite mass eigenvalues, $m_{i=1,2,3}$. The neutrino flavour states, $\nu_{l=e,\mu,\tau}$ are expressed as a linear superposition of the $\nu_{i}$ as shown below.

\begin{equation}
 \begin{bmatrix}
 \nu_{eL}\\
 \nu_{\mu L}\\
 \nu_{\tau L}\\
 \end{bmatrix}
 =
 U
 \begin{bmatrix}
 \nu_1\\
 \nu_2\\
 \nu_3\\
 \end{bmatrix},
 \end{equation}

where, 
\begin{equation}
U\,=\,\begin{bmatrix}
U_{e1} & U_{e2} & U_{e3}\\
U_{\mu1} & U_{\mu2} & U_{\mu3}\\
U_{\tau1} & U_{\tau2} & U_{\tau3}\\
\end{bmatrix}.
\end{equation}.

The $U$ is an unitary matrix and it physically signifies how the mass eigenstates mix to produce the flavour states. The matrix, U is known as Pontecorvo-Maki-Nakagawa-Sakata(PMNS) matrix and it is experimentally observable. As per the standard parametrization scheme adopted by Particle Data Group (PDG), the $U$ is represented as shown in the following,
 
\begin{equation}
U = \begin{bmatrix}
c_{12}c_{13} & c_{13}s_{12} & s_{13}e^{-i\delta}\\
-c_{23}s_{12}-c_{12}s_{13}s_{s23}e^{i\delta} & c_{12}c_{23}-s_{12}s_{13}s_{23}e^{i\delta} & c_{13}s_{23}\\
s_{12}s_{23}-c_{12}c_{23}s_{13}e^{i\delta} & -c_{12}s_{23}-s_{12}s_{13}c_{23}e^{i\delta} & c_{13}c_{23}\\
\end{bmatrix} \begin{bmatrix}
e^{i\alpha} & 0 & 0\\
0 & e^{i\beta} & 0\\
0 & 0 & 1\\
\end{bmatrix},
\end{equation}

where, $c_{ij}=\cos\theta_{ij}$, $s_{ij}=\sin\theta_{ij}$. The angles, $\theta_{12}$, $\theta_{23}$ and $\theta_{13}$ are termed as solar, atmospheric and reactor mixing angles. The phase angle, $\delta$ is called the Dirac CP violating phase; and $\alpha$ and $\beta$ are called the Majorana phases.

For two neutrino scenario, the neutrino mixing is depicted as shown in the following

\begin{equation}
 \begin{bmatrix}
 \nu_{e}\\
 \nu_{\mu}\\
 \end{bmatrix}
 =
 \begin{bmatrix}
 \cos\theta & \sin\theta\\
 -\,\sin\theta & \cos\theta\\
 \end{bmatrix}
 \begin{bmatrix}
 \nu_1\\
 \nu_2\\
 \end{bmatrix},
 \label{two flavour mixing}
 \end{equation}
where, $\theta$ is the mixing angle.

\section{Neutrino oscillation Probability Inside an Infinite Square Well Potential \label{section 3}}

The conversion from one flavour to another in the process of neutrino oscillation is studied in terms of the Probability $P_{\nu_e\rightarrow\nu_{\mu}}$ which from Pontecorvo's original formulation can be expressed as,

\begin{equation}
 P_{\nu_e\rightarrow\nu_{\mu}}\,=\,\sin^2(2\,\theta)\,\sin^2 \Bigg(\frac{\Delta m_{21}^2c^3 l}{4 E \hslash}\Bigg),
 \label{2x2 probability expression}
 \end{equation}

where, $\Delta m_{21}^2\,=\,m^2_2-m^2_1$ and the angle $\theta$ is the mixing angle. The $l$ is the distance between source and detector and $E$ stands for the energy of the neutrinos. This is to be noted that the above relation is true for two neutrino flavour scenario, and while deriving the same the neutrinos are considered as free ultra relativistic particles. But in connection with the present work, we shall try to look into the possibilities of neutrino oscillation inside an one dimensional infinite square-well potential. In general, the form of a infinite square well potential is defined in the following way,

\begin{equation}
V(x)=\Bigg\{\begin{matrix}
0 & \,\,\,\,0<x<L\\\\
\infty & \,\,\,\, \text{otherwise}
\end{matrix},
\end{equation}

where, L is the length of the box. Whether the particle in a box behaves as relativistic or non-relativistic depends upon the parameter $L_c\,=\,\frac{\lambda_c}{2\,\pi}$, where, $\lambda_c$ is the Compton wavelength of the particle inside the well. The $\lambda_c$ can be expressed as shown below,

\begin{equation}
\lambda_c=\frac{h}{m c},
\label{compton wavelength}
\end{equation}

  where, $m$ is the mass of the particle. Depending upon the condition whether $L>>L_c$ or $L \lesssim L_c$, we identify the particle inside the box as non-relativistic and relativistic respectively \cite{Alberto_1996}. Keeping aside the spin $1/2$ nature of neutrino, we employ both Schr$\ddot{o}$dinger equation (for non-relativistic case) and Klein-Gordon equation (for relativistic case) respectively to study the probability of flavour conversion. The same is studied in the framework of both Dirac and Majorana equation which respects the spin $1/2$ nature of neutrino. 

 In the present work, we stick to the normal ordering of the neutrino masses ($m_3>m_2>m_1$). The value of $m_3$ is fixed at $0.06\,eV/c^2$ and $m_1$ and $m_2$ are calculated from two mass squared differences ($\Delta\,m_{21}^2$ and $\Delta\,m_{31}^2$) as $0.0318\,eV/c^2$ and $0.0329\,eV/c^2$. We highlight that the sum of three neutrino masses ($m_1\,+\,m_2\,+\,m_3\,=\,0.124\,eV/c^2$) considered in our work is consistent with the cosmological data \cite{Planck:2018vyg}. In SI unit system, the mass eigenvalues are of the order of $10^{-38} kg$. From Eq.\,(\ref{compton wavelength}), the Compton wavelength associated with the mass eigenvalues comes out to be $\approx10^{-5}\,m$. For the non-relativistic case, we set the well length $L$ at $1$ meter which is $>>L_c$ while for relativistic scenario, the well length is set at $10^{-5}$ meter which is of the order of the Compton wavelength of neutrino.

We solve the Schr$\ddot{o}$dinger equation for the well and obtain the mass eigenstates as shown in the following,

\begin{eqnarray}
\nu_1\,&=&\,\sqrt{\frac{2}{L}}\,\sin\Big(\frac{n_1 \pi x}{L}\Big) e^{-\frac{i}{\hslash} E_{n_1}t} \label{mass eigenstate 1},\\
\nu_2\,&=&\,\sqrt{\frac{2}{L}}\,\sin\Big(\frac{n_2 \pi x}{L}\Big)e^{-\frac{i}{\hslash}E_{n_2}t},
\label{mass eigenstate 2}
\end{eqnarray}

where, $n_i$ is the quantum number associated with the allowed energy states.  In the present work, the quantum numbers $n_1$ and $n_2$ are coupled to the neutrino mass eigenstates $\nu_1$ and $\nu_2$ respectively. The quantised energy of the particle in $n^{th}_i$ state inside the potential is given by,

\begin{equation}
E_{n_i} = \frac{n^2_i\,\pi^2\,\hslash^2}{2\,m_i\,L^2},
\label{quantised energy}
\end{equation}

where, $m_i$ is the mass of the particle. 

We find that the particle may lie in any one of the quantized energy states depicted by the quantum number $n$. In this regard, we expect that the oscillation probability will depend on the quantum number of the corresponding mass eigenstates.

In our case, an electron type neutrino is trapped inside a hypothetical infinite square well potential. We first consider two flavour oscillation and derive the expression for $P_{\nu_e\rightarrow\nu_{\mu}}$ (see \ref{appendix 1}) as shown in the following,

\begin{equation}
P_{\nu_e\rightarrow\nu_{\mu}}\,=\,\sin^2(2\,\theta)\,\sin^2 \Big[\frac{\pi^2\,\hslash\,t}{4\,L^2}\Big(\frac{n_1^2}{m_1}-\frac{n_2^2}{m_2}\Big)\Big].
\label{2x2 probability}
\end{equation}

It is seen that not only the masses of corresponding mass eigenstates of $\nu_1$ and $\nu_2$, but also the energy states where the mass eigenstates lie will contribute to the oscillation probability.

Similarly, for three neutrino case, the expression of $P_{\nu_e\rightarrow\nu_\mu}$ can be derived as shown below,

\begin{eqnarray}
P_{\nu_e\rightarrow\nu_\mu}&=&-4\,Re\,(U_{e1}U^*_{\mu1}U^*_{e2}U_{\mu2})\,\sin^2\Big[\frac{\pi^2 \hslash t}{4L^2}\Big(\frac{n^2_2}{m_2}-\frac{n^2_1}{m_1}\Big)\Big]+2\,Im\,(U_{e1}U^*_{\mu1}U^*_{e2}U_{\mu2})\,\sin\Big[ \frac{\pi^2 \hslash t}{2L^2}\Big(\frac{n^2_2}{m_2}-\,\nonumber\\&&\frac{n^2_1}{m_1}\Big)\Big]\,-4\,Re\,(U_{e1}U^*_{\mu1}U^*_{e3}U_{\mu3})\,\sin^2\Big[\frac{\pi^2 \hslash t}{4L^2}\Big(\frac{n^2_3}{m_3}-\frac{n^2_1}{m_1}\Big)\Big]+2\,Im\,(U_{e1}U^*_{\mu1}U^*_{e3}U_{\mu3})\,\sin\Big[ \frac{\pi^2 \hslash t}{2L^2}\nonumber\\&&\,\Big(\frac{n^2_3}{m_3}-\frac{n^2_1}{m_1}\Big)\Big]\,-\,4\,Re\,(U_{e2}U^*_{\mu2}U^*_{e3}U_{\mu3})\,\sin^2\Big[\frac{\pi^2 \hslash t}{4L^2}\Big(\frac{n^2_3}{m_3}-\frac{n^2_2}{m_2}\Big)\Big]+2\,Im\,(U_{e2}U^*_{\mu2}U^*_{e3}U_{\mu3})\,\nonumber\\&&\sin\Big[ \frac{\pi^2 \hslash t}{2L^2}\Big(\frac{n^2_3}{m_3}-\frac{n^2_2}{m_2}\Big)\Big].
\label{3x3 probability}
\end{eqnarray}

For relativistic case, dealing with the problems related to trapping potential is challenging because one may encounter a conjecture termed as Klein Paradox. Precisely, the reflected flux from the wall is greater than that of the incident flux. However, we may avoid this problem and rescue the one-particle picture by adopting the concept of position-dependent mass \cite{Alberto:2011wn, Alberto:2017pkj},

\begin{equation}
m(x)=\Bigg\{\begin{matrix}
m & \,\,\,\,0<x<L\\\\
\infty & \,\,\,\, \text{otherwise}
\end{matrix}.
\end{equation}

We solve Klein-Gordon equation to obtain the mass eigenstates. We see that the states are identical to those appearing in Eqs.\,(\ref{mass eigenstate 1}-\ref{mass eigenstate 2})provided the expression of quantised energy, $E_{n_i}$ of the mass eigenstate $\nu_i$ is given by,

\begin{equation}
E_{n_{i}}\,=\,\sqrt{\frac{n_{i}^2 \pi^2 \hslash^2 c^2}{L^2}\,+\,m_i^2c^4},
\end{equation}

where, $i\,=\,1,2$. Now, if we consider a relativistic electron type neutrino trapped inside an infinite square-well potential, the expression for $P_{\nu_e\rightarrow\nu_\mu}$ (in case of two flavour oscillation) can be derived in the following manner,

\begin{eqnarray}
P_{\nu_e\rightarrow\nu_\mu}\,&=&\,\sin^2({2\theta})\,\sin^2 \Big[\frac{t}{2\hslash}\Big\{\Big(\frac{n^2_1\pi^2 \hslash^2 c^2}{L^2}\,+\,m^2_1c^4\Big)^{\frac{1}{2}}-\Big(\frac{n^2_2 \pi^2 \hslash^2 c^2}{L^2}\,+\,m^2_2c^4\Big)^{\frac{1}{2}}\Big\}\,\,\Big].
\label{2x2 probability for relativistic case}
\end{eqnarray}

If we take the non-relativistic limit on Eq.\,(\ref{2x2 probability for relativistic case}), the expression of $P_{\nu_e\rightarrow\nu_\mu}$ is obtained as shown below,

\begin{eqnarray}
P_{\nu_e\rightarrow\nu_\mu}^{NRL(KG)}\,&=&\,\sin^2(2\,\theta)\,\sin^2 \Big[\frac{\pi^2\,\hslash\,t}{4\,L^2}\Big(\frac{n_1^2}{m_1}-\frac{n_2^2}{m_2}\Big)\Big].
\label{2x2 probability for KG in non relativistic limit}
\end{eqnarray}

It is evident from Eq.\,(\ref{2x2 probability for KG in non relativistic limit}), under non-relativistic limit, the $P_{\nu_e\rightarrow\nu_\mu}$ becomes identical to the Schr$\ddot{o}$dinger case. On the other hand, under the ultra-relativistic limit, the Eq.\,(\ref{2x2 probability for relativistic case}) take the following form,

\begin{eqnarray}
P_{\nu_e\rightarrow\nu_\mu}^{URL(KG)}\,&=&\,\sin^2({2\theta})\,\sin^2 \Big[\frac{t}{2\hslash}\Big\{\frac{h \pi c}{L}(n_1-n_2)+\frac{c^3 L}{2 \pi \hslash}\Big(\frac{m_1^2}{n_1}-\frac{m_2^2}{n_2}\Big)\Big\}\,\,\Big].
\label{2x2 probability for KG in relativistic limit}
\end{eqnarray}

Now, if we consider the spin half nature of the neutrino, the neutrino oscillation can be studied in the light of Dirac equation. We solve the Dirac equation inside the well and obtain the following form of the mass eigenstates\cite{Alberto_1996},

\begin{eqnarray}
\nu_i &=& B e^{i\frac{k_i L}{2}}\begin{pmatrix}
2\cos\big(k_i x-\frac{k_i L}{2}\big)\chi\\
2i\,r \sin\big(k_i x-\frac{k_i L}{2}\big)\sigma_x\chi
\end{pmatrix} e^{-\frac{i}{\hslash}E_i t}, \quad \quad i=1,2.
\end{eqnarray}

where, $\chi$ is either $(1,0)^T$ or $(0,1)^T$ and $r=\frac{c k_i \hslash}{E+mc^2}$. The Pauli spin matrix, $\sigma_x$ is chosen in a basis where it is diagonal(This is in agreement with the fact that the movement of neutrino is constrained in the $x$ direction). Here, $k_i$ stands for the wave number associated with $\nu_i$. However, $k_i$ takes discrete values which can be found out by solving the following transcendental equation,

\begin{equation}
\tan(k_i L)\,=\,-\,\frac{\hslash k_i}{m_i c},
\label{transcendental equation}
\end{equation}

and further, the discrete energy eigenvalues $E_i$ of the state $\nu_{i}$ are worked out using the following relation,

\begin{equation}
E_i\,=\,\sqrt{\hslash^2 c^2 k^2_i\,+\,m_i^2c^4}, \,\,\, k_i\,\neq\,0.
\label{Dirac Energy}
\end{equation}

Now, if we consider two neutrino case, the expression for $P_{\nu_e\rightarrow\nu_\mu}$ can be obtained as shown below,

\begin{eqnarray}
P_{\nu_e\rightarrow\nu_\mu}\,&=&\,\sin^2({2\theta})\,\sin^2 \Big[\frac{t}{2\hslash}\Big\{\Big(\hslash^2 c^2 k^2_1\,+\,m^2_1 c^4\Big)^{\frac{1}{2}}\,-\,\Big(\hslash^2 c^2 k^2_2\,+\,m^2_2 c^4\Big)^{\frac{1}{2}}\Big\}\,\,\Big].
\label{2x2 probability for Dirac case}
\end{eqnarray}

Under the non-relativistic limit, Eq.\,(\ref{2x2 probability for Dirac case}) can be expressed as,

\begin{eqnarray}
P_{\nu_e\rightarrow\nu_\mu}^{NRL(Dirac)}\,&=&\,\sin^2({2\theta})\,\sin^2 \Big[\frac{ \hslash t}{4}\Big(\frac{k_1^2}{m_1}-\frac{k_2^2}{m_2}\Big)\,\Big],
\label{2x2 probability for Dirac case in non-relativistic limit}
\end{eqnarray}

whereas, for the ultra-relativistic limit, the Eq.\,(\ref{2x2 probability for Dirac case})appears as shown below,

\begin{eqnarray}
P_{\nu_e\rightarrow\nu_\mu}^{URL(Dirac)}\,&=&\,\sin^2({2\theta})\,\sin^2 \Big[\frac{t}{2\hslash}\Big\{\hslash c\,(k_1-k_2)+\frac{c^3}{2\hslash}\Big(\frac{m_1^2}{k_1}-\frac{m_2^2}{k_2}\Big)\Big\}\,\Big],
\label{2x2 probability for Dirac case in relativistic limit}
\end{eqnarray}

Similarly, for three neutrino case, we can derive the expression of $P_{\nu_e\rightarrow\nu_\mu}$ as shown below,

\begin{eqnarray}
P_{\nu_e\rightarrow\nu_\mu} &=&\,-\,4\,Re\,(U_{e1}U^*_{\mu1}U^*_{e2}U_{\mu2})\,\sin^2\,\Big[\frac{t}{2\hslash}\Big\{\big(k^2_{2}\hslash^2c^2\,+\,m^2_2c^4\big)^{\frac{1}{2}}-\big(k^2_{1}\hslash^2c^2\,+\,m^2_1c^4\big)^{\frac{1}{2}}\Big\}\Big]\nonumber\\&&+\,2\,Im\,(U_{e1}\,U^*_{\mu1}U^*_{e2}U_{\mu2})\,\sin\,\Big[\frac{t}{\hslash}\Big\{\big(k^2_{2}\hslash^2c^2\,+\,m^2_2c^4\big)^{\frac{1}{2}}-\big(k^2_{1}\hslash^2c^2\,+\,m^2_1c^4\big)^{\frac{1}{2}}\Big\}\Big]\,\nonumber\\&&\,-\,4\,Re\,(U_{e1}U^*_{\mu1}U^*_{e3}U_{\mu3})\,\sin^2\,\Big[\frac{t}{2\hslash}\Big\{\big(k^2_{3}\hslash^2c^2\,+\,m^2_3c^4\big)^{\frac{1}{2}}-\big(k^2_{1}\hslash^2c^2\,+\,m^2_1c^4\big)^{\frac{1}{2}}\Big\}\Big]\,\nonumber\\&&+\,2\,Im\,(U_{e1}\,U^*_{\mu1}U^*_{e2}U_{\mu2})\,\sin\, \Big[\frac{t}{\hslash}\Big\{\big(k^2_{3}\hslash^2c^2\,+\,m^2_3c^4\big)^{\frac{1}{2}}-\big(k^2_{1}\hslash^2c^2\,+\,m^2_1c^4\big)^{\frac{1}{2}}\Big\}\Big]\,\nonumber\\&&\,-\,4\,Re\,(U_{e2}U^*_{\mu2}U^*_{e3}U_{\mu3})\,\sin^2\,\Big[\frac{t}{2\hslash}\Big\{\big(k^2_{3}\hslash^2c^2\,+\,m^2_3c^4\big)^{\frac{1}{2}}-\big(k^2_{2}\hslash^2c^2\,+\,m^2_2c^4\big)^{\frac{1}{2}}\Big\}\Big]\nonumber\\&&+\,2\,Im\,(U_{e2}U^*_{\mu2}U^*_{e3}U_{\mu3})\,\sin\, \Big[\frac{t}{\hslash}\Big\{\big(k^2_{3}\hslash^2c^2\,+\,m^2_3c^4\big)^{\frac{1}{2}}-\big(k^2_{2}\hslash^2c^2\,+\,m^2_2c^4\big)^{\frac{1}{2}}\Big\}\Big].\,\nonumber\\&&
\label{3x3 probability for Dirac case}
\end{eqnarray}

Here, we wish to highlight that the above equation is true for the Klein Gordon scenario as well. But in contrast, for Klein Gordon case, $k_{i=1,2,3}$ appearing in above equation is equivalent to $k_{n_{i=1,2,3}}$, where, $k_{n_{i}}= \frac{n_{i} \pi}{L}$.

\section{Important Observations \label{section 4}}

As it is discussed in section\,\ref{section 3}, in addition to the mass eigenvalues, the quantum number $n$ also contribute to the neutrino oscillation probability. In this connection, we can expect that the frequency of oscillation ($f$) will also vary with the energy (i.e., quantum number $n$) of the mass eigenstates. In this regard, we concentrate on two cases:

\begin{enumerate}
\item [(i)] \textbf{case-I}:\,The neutrino mass eigenstates carry same quantum number (For two neutrino scenario, $n=n_1=n_2$ and the same for three neutrino case refers to $n=n_1=n_2=n_3$).
\item [(ii)] \textbf{case-II}: \,The neutrino mass eigenstates bear different quantum numbers. 
\end{enumerate}

Needless to mention that the situation $n_1=n_2$ does not imply that the two neutrinos lie in the same energy eigenstate because they differ in masses. The variation of the oscillation frequency with respect to the quantum number $n$ can be visualised from the parameter $f'$, \footnote{In the present work, though the quantum number $n$ is a desecrate variable, the $f'$ is obtained to understand the variation of the oscillation frequency with respect to the quantum number $n$.} where, $f'=\frac{df}{dn}$.

For the non-relativistic scenario, the expression of oscillation frequency $f_{NR}$ can be derived in the following manner,

\begin{equation}
f_{NR}\,=\,\frac{\pi\,\hslash}{8\,L^2}\Big(\frac{n_1^2}{m_1}-\frac{n_2^2}{m_2}\Big).
\end{equation}

\begin{center}
\begin{figure}[!]
   \begin{center}
    \subfigure[]{\includegraphics[width=0.24\textwidth]{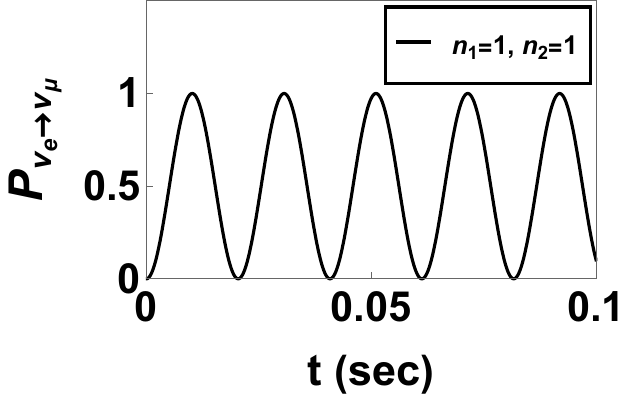}} 
    \subfigure[]{\includegraphics[width=0.24\textwidth]{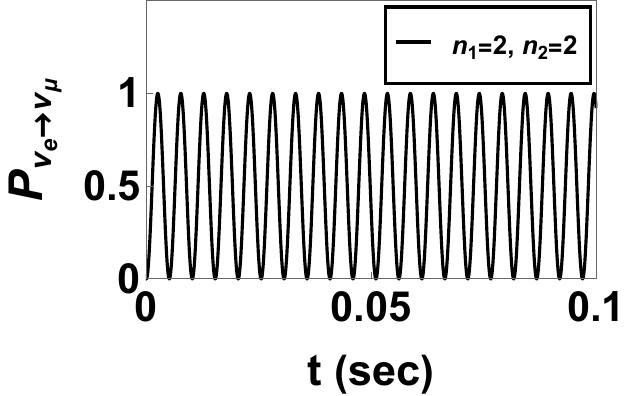}}
    \subfigure[]{\includegraphics[width=0.24\textwidth]{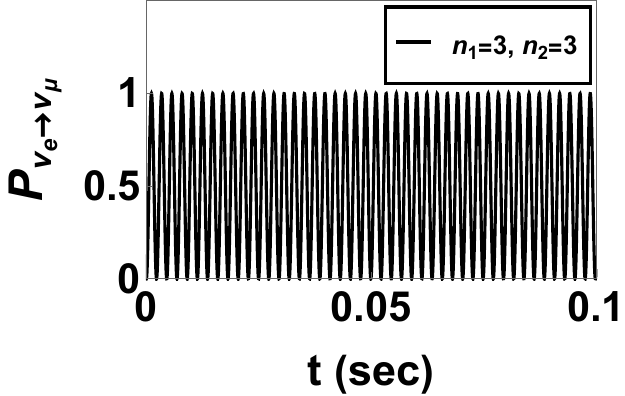}} 
   \end{center}
 \caption{(a) Shows the plot of $P_{\nu_e\rightarrow\nu_{\mu}}$ against $t$ for $n_1=1$ and $n_2=1$ in case of two flavor oscillation in non-relativistic case. (b) Plot of $P_{\nu_e\rightarrow\nu_{\mu}}$ against $t$ is shown when $n_1=2$ and $n_2=2$. (c) Presents the plot of $P_{\nu_e\rightarrow\nu_{\mu}}$ against $t$ for $n_1=3$ and $n_2=3$. From the plots, we notice that the oscillation frequency increases as the energy of the mass eigenstates increases.}
\label{fig:2x2 non relativistic}
\end{figure}
\end{center}

\begin{center}
\begin{figure}[!]
   \begin{center}
     \subfigure[]{\includegraphics[width=0.24\textwidth]{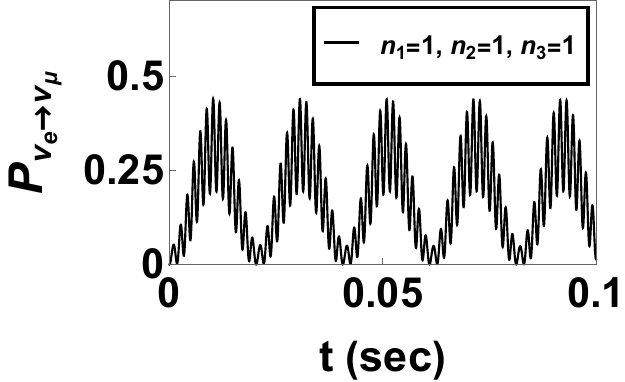}} 
    \subfigure[]{\includegraphics[width=0.24\textwidth]{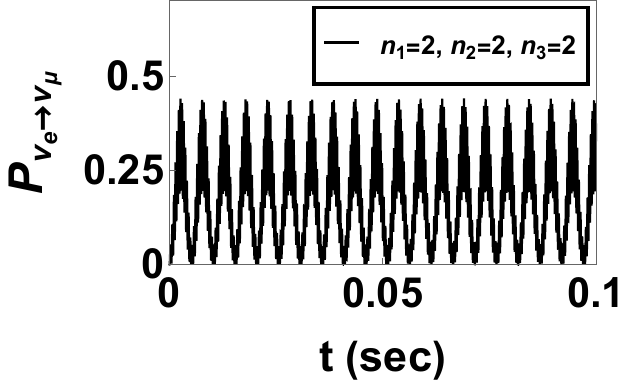}}
    \subfigure[]{\includegraphics[width=0.24\textwidth]{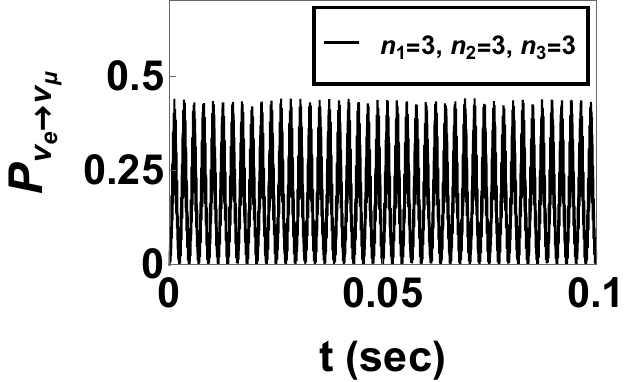}} 
    \end{center} 
    \caption{(a) Displays the plot of $P_{\nu_e\rightarrow\nu_{\mu}}$ against $t$ when $n_1=1$, $n_2=1$ and $n_3=1$ for three flavor oscillation in non-relativistic scenario . (b) Plot of $P_{\nu_e\rightarrow\nu_{\mu}}$ against $t$ is shown where $n_1=2$, $n_2=2$ and $n_3=2$. (c) Gives the plot of $P_{\nu_e\rightarrow\nu_{\mu}}$ against $t$ for $n_1=3$, $n_2=3$ and $n_3=3$. From the plots, we see a increase in oscillation frequency with the increase in energy of the mass eigenstates.}
\label{fig:3x3 non relativistic}
\end{figure}
\end{center}

For the \textbf{case-I}, it is seen that the oscillation frequency increases if the mass eigenstates take higher values $n$\,(see Figs.\,\ref{fig:2x2 non relativistic}-\ref{fig:3x3 non relativistic}). The variation of the frequency with respect to the quantum number $n$ is given by,

\begin{equation}
f'=\frac{n \pi \hslash}{4 L^2}\Big(\frac{1}{m_1}-\frac{1}{m_2}\Big).
\label{variation of oscillation frequency in non-relativistic case}
\end{equation} 

\begin{figure}[!]
\centering
    \subfigure[]{\includegraphics[width=0.24\textwidth]{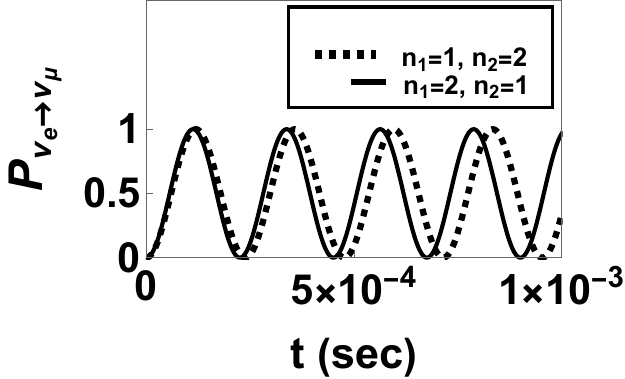}} 
    \subfigure[]{\includegraphics[width=0.24\textwidth]{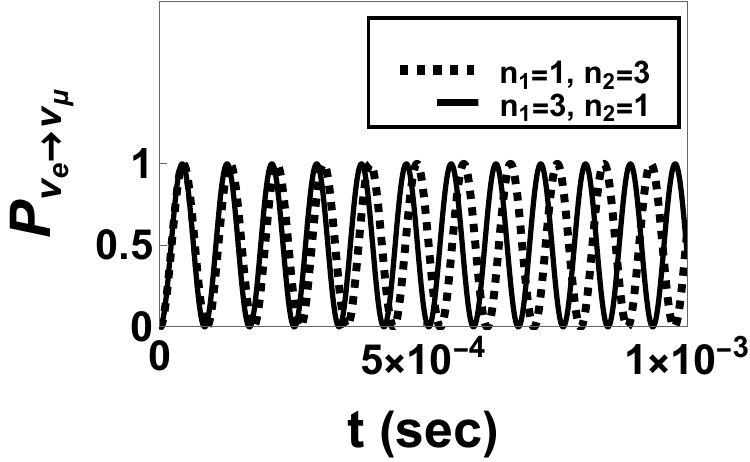}}
    \subfigure[]{\includegraphics[width=0.24\textwidth]{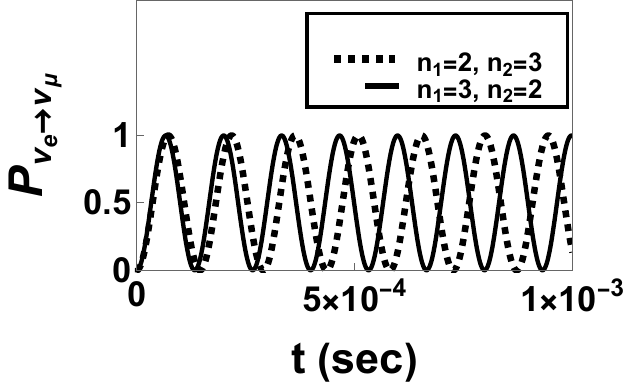}}
 \caption{(a) Presents the plot of $P_{\nu_e\rightarrow\nu_{\mu}}$ against $t$ for $n_1=1$ and $n_2=2$ and vice-versa for two flavor oscillation in non-relativistic case. (b) Plot of $P_{\nu_e\rightarrow\nu_{\mu}}$ against $t$ is shown for $n_1=1$ and $n_2=3$ and vice-versa. (c) Manifests the plot of $P_{\nu_e\rightarrow\nu_{\mu}}$ against $t$ for $n_1=2$ and $n_2=3$ and vice-versa. From the plots, we find that the oscillation frequency changes with the change in energy of the mass eigenstates.}
\label{fig:2x2 non relativistic for differnt energy}
\end{figure}

\begin{figure}[!]
\centering
    \subfigure[]{\includegraphics[width=0.25\textwidth]{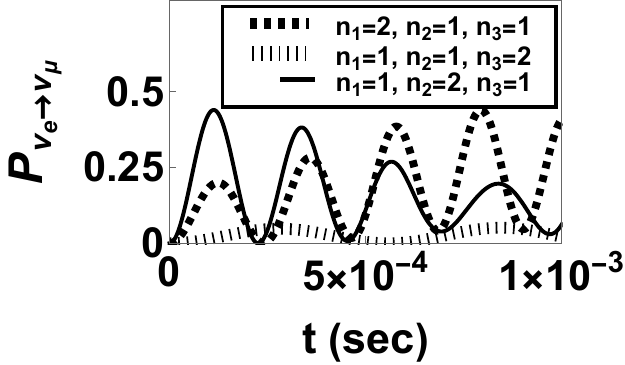}} 
    \subfigure[]{\includegraphics[width=0.25\textwidth]{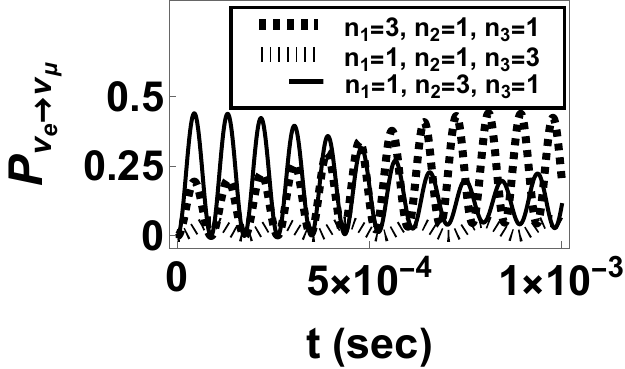}}
    \subfigure[]{\includegraphics[width=0.25\textwidth]{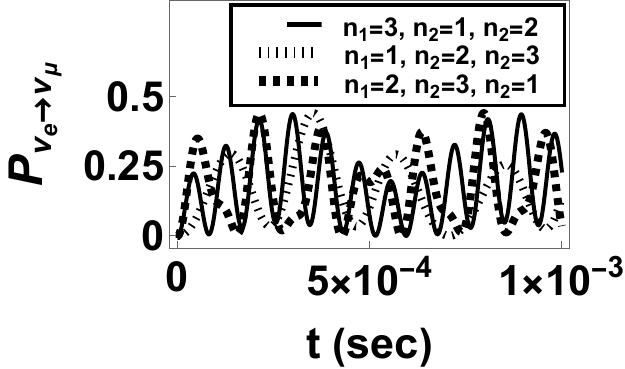}}
 \caption{(a) Shows the plot of $P_{\nu_e\rightarrow\nu_{\mu}}$ against $t$ for three flavor oscillation in non-relativistic case, where, $n_1$, $n_2$ and $n_3$ can have values $(1,1,2)$, $(1,2,1)$ and $(2,1,1)$ respectively. (b) Gives the plot of $P_{\nu_e\rightarrow\nu_{\mu}}$ against $t$ for the combinations of quantum numbers $(1,1,3)$, $(1,3,1)$ and $(3,1,1)$ respectively. (c) Plot of $P_{\nu_e\rightarrow\nu_{\mu}}$ against $t$ is shown  where $n_1$, $n_2$ and $n_3$ are chosen to be $(1,2,3)$, $(2,3,1)$ and $(3,1,2)$ respectively. From the plots, it is clear that the  frequency as well the amplitude of the oscillation changes as the energy of the mass eigenstates changes.}
\label{fig:3x3 non relativistic for differnt energy}
\end{figure}

From Eq.\,(\ref{variation of oscillation frequency in non-relativistic case}), we see that the change in oscillation frequency is directly proportional to the quantum number $n$. So it is obvious that the frequency will increase with the increase in the quantum number. For the \textbf{case-II}, we notice that oscillation frequency as well as the amplitude of oscillation varies when the energy of the mass eigenstates varies(see Figs.\,\ref{fig:2x2 non relativistic for differnt energy}-\ref{fig:3x3 non relativistic for differnt energy}). This is because altering the energy states will certainly shift the frequency as the masses corresponding to the mass eigenstates are different.

Similar to the non-relativistic case, we emphasize over two possible scenarios. The oscillation frequency for Klein-Gordon case $f_{KG}$ is given by,

\begin{eqnarray}
f_{KG}\,&=&\,\frac{1}{4\pi\hslash}\Bigg[\Big(\frac{n^2_1\pi^2 \hslash^2 c^2}{L^2}\,+\,m^2_1c^4\Big)^{\frac{1}{2}}-\Big(\frac{n^2_2 \pi^2 \hslash^2 c^2}{L^2}\,+\,m^2_2c^4\Big)^{\frac{1}{2}}\Bigg].
\label{KG Frequency}
\end{eqnarray}

\begin{figure}[!]
    \begin{center}
    \subfigure[]{\includegraphics[width=0.24\textwidth]{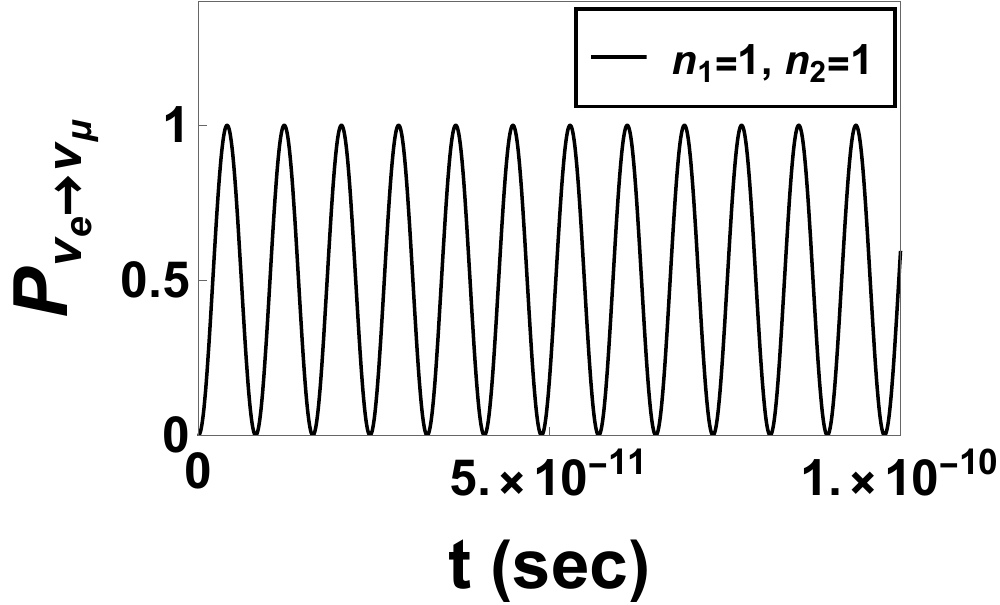}} 
    \subfigure[]{\includegraphics[width=0.24\textwidth]{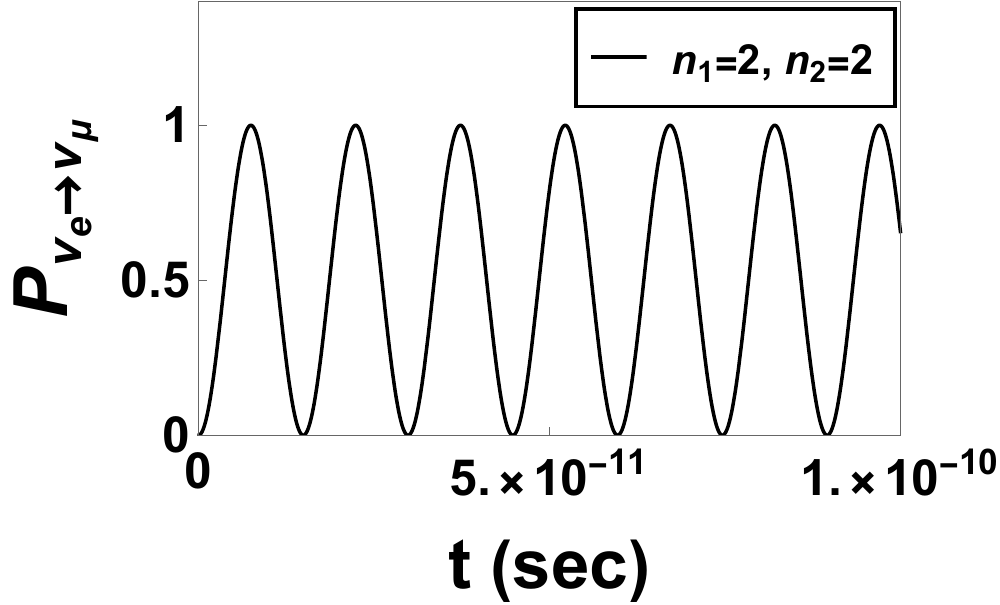}}
    \subfigure[]{\includegraphics[width=0.24\textwidth]{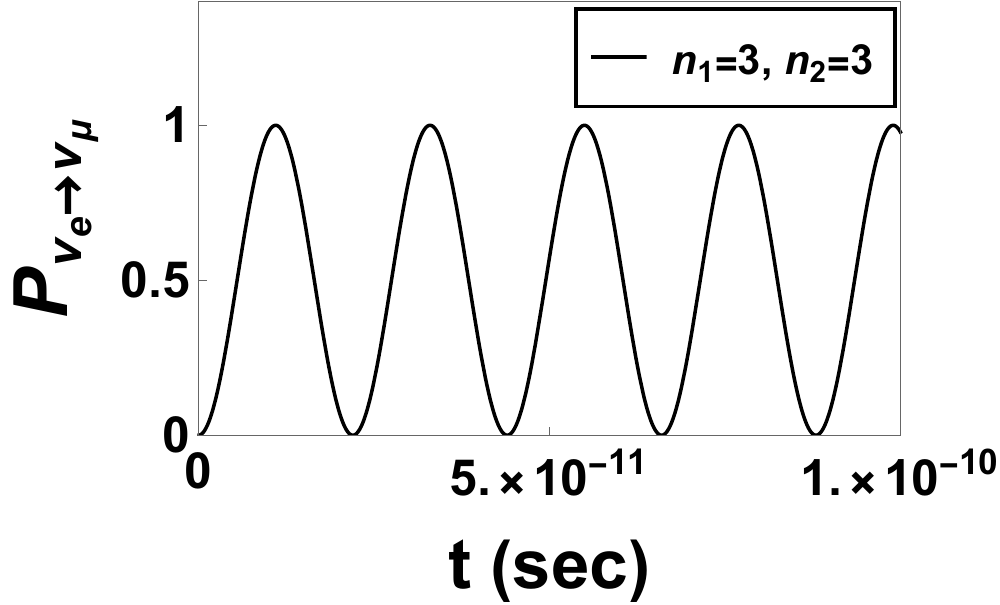}}
    \end{center}
    \caption{(a) Shows the plot of $P_{\nu_e\rightarrow\nu_{\mu}}$ against $t$ where, $n_1=1$ and $n_2=1$ for two flavor oscillation in relativistic case (Klein-Gordon). (b) The plot of $P_{\nu_e\rightarrow\nu_{\mu}}$ against $t$ is manifested for $n_1=2$ and $n_2=2$. (c) Displays the plot of $P_{\nu_e\rightarrow\nu_{\mu}}$ against $t$ when $n_1=3$ and $n_2=3$. From the plots, we notice a decrease in the oscillation frequency as the energy of the mass eigenstates increases.}
\label{fig:2x2 relativistic}
\end{figure}

\begin{figure}[!]
    \begin{center}
     \subfigure[]{\includegraphics[width=0.24\textwidth]{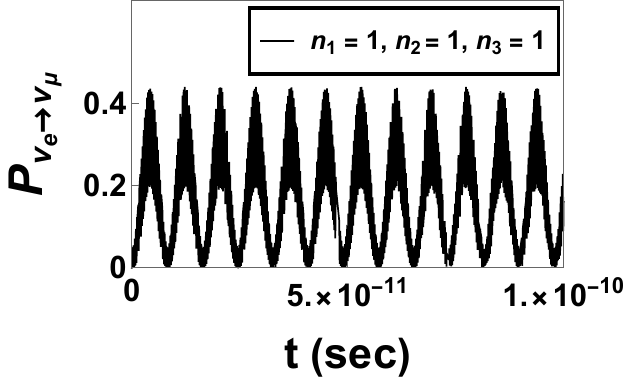}} 
    \subfigure[]{\includegraphics[width=0.24\textwidth]{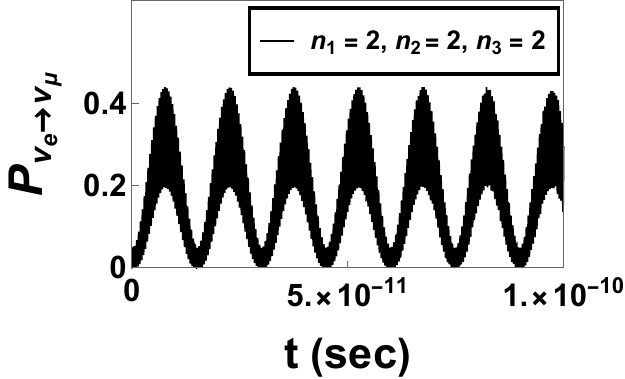}}
    \subfigure[]{\includegraphics[width=0.24\textwidth]{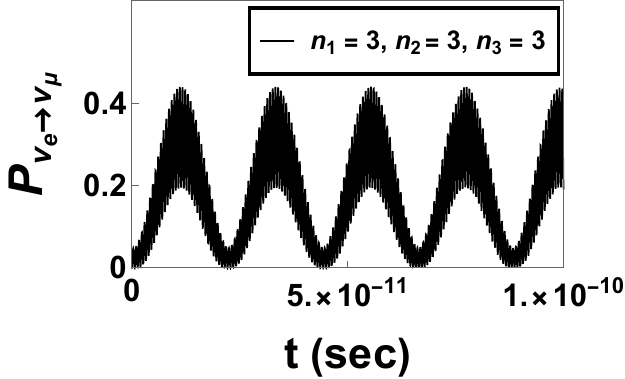}} 
     \end{center} 
    \caption{(a) Shows the plot of $P_{\nu_e\rightarrow\nu_{\mu}}$ against $t$, where, $n_1=1$, $n_2=1$ and $n_3=1$ for three flavor oscillation in relativistic scenario (Klein-Gordon). (b) The plot of $P_{\nu_e\rightarrow\nu_{\mu}}$ against $t$ is manifested for $n_1=2$, $n_2=2$ and $n_3=2$. (c) Presents the plot of $P_{\nu_e\rightarrow\nu_{\mu}}$ against $t$ for $n_1=3$, $n_2=3$ and $n_3=3$ respectively. From the plots, we observe that the oscillation  frequency decreases with the increase in energy of the mass eigenstates.}
\label{fig:3x3 relativistic}
\end{figure}

For the \textbf{case-I}, it is found that the frequency of oscillation decreases as the energy of the mass eigenstates increases (see Figs.\,\ref{fig:2x2 relativistic}-\ref{fig:3x3 relativistic}). In the relativistic limit, the variation of the frequency with respect to the quantum number $n$ is given by,

\begin{equation}
f' = \frac{L c^3}{8 \pi^2 \hslash^2 n^2}\,(m_2^2 - m_1^2).
\label{variation of KG oscillation frequency in relativistic case limit}
\end{equation}

It is clearly seen from Eq.\,(\ref{variation of KG oscillation frequency in relativistic case limit}) that the quantity $f'$ is inversely proportional to square of the quantum number $n$. Under non-relativistic limit of the probability corresponding to the Klein-Gordon case, the expression of frequency is given by,

\begin{equation}
f_{NRL}=\frac{1}{4 \pi \hslash}\Big[\frac{\pi^2 \hslash^2}{2 L^2}\Big(\frac{n_1^2}{m_1}-\frac{n_2^2}{m_2}\Big)\Big]
\end{equation}

The variation of frequency with respect to the quantum number $n$ can be visualised as shown in the following,

\begin{equation}
f_{NRL}'=\frac{n \pi \hslash}{4 L^2}\Big(\frac{1}{m_1}-\frac{1}{m_2}\Big).
\end{equation}

It is to be noted the $f_{NRL}'$ is identical to Eq.\,(\ref{variation of oscillation frequency in non-relativistic case}). Similarly, for the Dirac case, the oscillation frequency $f_D$ can be calculated from the expression as shown in the following,

\begin{eqnarray}
f_D\,&=&\,\frac{1}{4\pi\hslash}\Bigg[\Big(\hslash^2 c^2 k^2_1\,+\,m^2_1 c^4\Big)^{\frac{1}{2}}\,-\,\Big(\hslash^2 c^2 k^2_2\,+\,m^2_2 c^4\Big)^{\frac{1}{2}}\Bigg].
\end{eqnarray}

\begin{figure}[!]
\centering
    \subfigure[]{\includegraphics[width=0.22\textwidth]{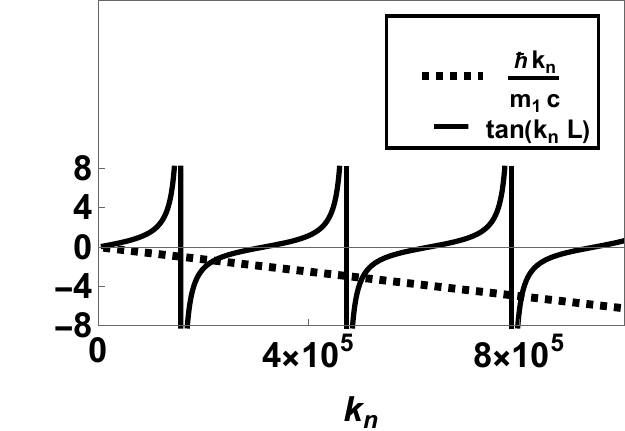}} 
    \subfigure[]{\includegraphics[width=0.22\textwidth]{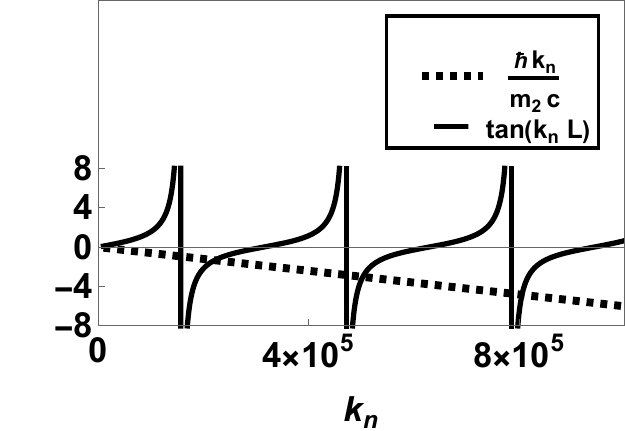}}
    \subfigure[]{\includegraphics[width=0.22\textwidth]{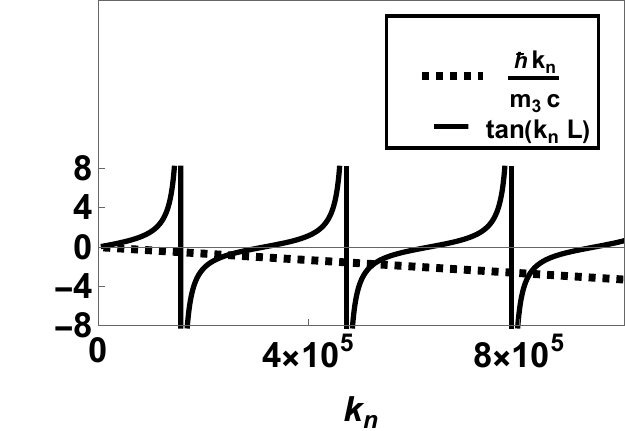}}  
    \caption{(a) Presents the plot of $\tan(kL)$ and $\frac{-\hslash k}{m_1 c}$ against the wave number $k$. (b) The plot of $\tan(kL)$ and $\frac{-\hslash k}{m_2 c}$ against the wave number $k$ is displayed. (c) Manifests the plot of $\tan(kL)$ and $\frac{-\hslash k}{m_3 c}$ against the wave number $k$. From figure, we encounter three intersection points of the above said expressions. From these points, the values of $k_1$, $k_2$ and $k_3$ are calculated for three mass eigenstates.}
\label{fig:transcendental equation}
\end{figure}

In the above expression, the discrete values of $k_{i=1,2}$ are estimated using Eq.(\ref{transcendental equation}). For this we first plot the left hand side and right hand side of the said equation against $k_i$ \ref{fig:transcendental equation}) and estimate the points where the plots intersect and then using the Newton-Raphson method, we determine the possible values of  $k_i$ for different mass eigenstates \,(see Table \ref{table:Numerical values}).

\begin{figure}[!]
\centering
    \subfigure[]{\includegraphics[width=0.24\textwidth]{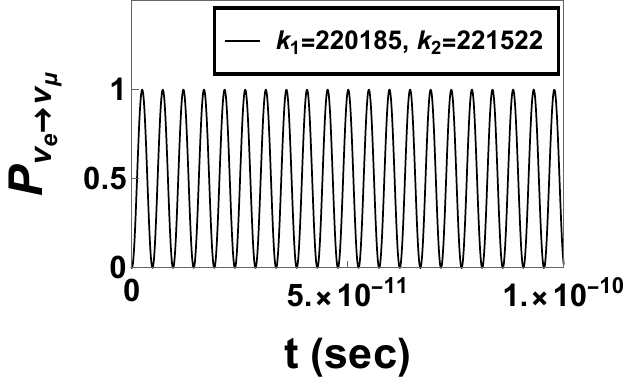}} 
    \subfigure[]{\includegraphics[width=0.24\textwidth]{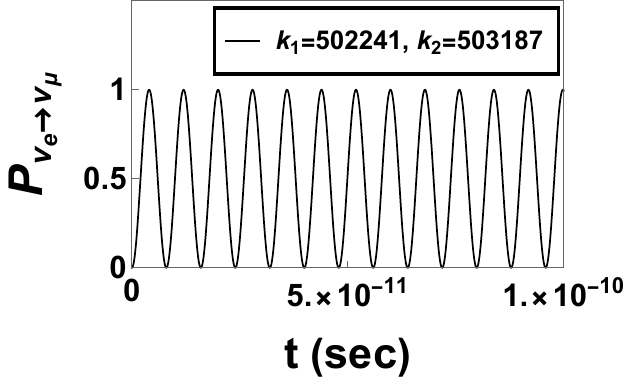}}
    \subfigure[]{\includegraphics[width=0.24\textwidth]{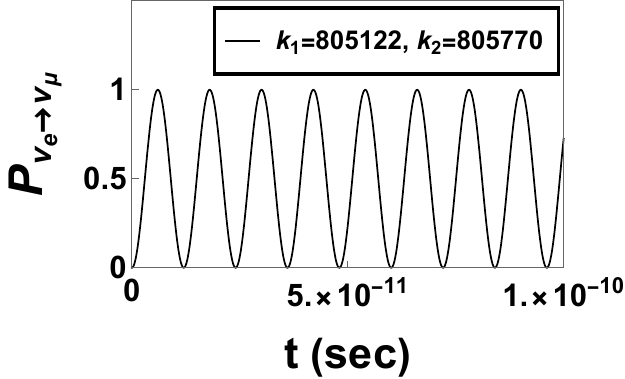}}  
 \caption{(a) Displays the plot of $P_{\nu_e\rightarrow\nu_{\mu}}$ against $t$ where, $k_1=220185\,m^{-1}$ and $k_2=221522\,m^{-1}$ for two flavor oscillation in relativistic case (Dirac). (b) Presents the plot of $P_{\nu_e\rightarrow\nu_{\mu}}$ against $t$ for $k_1=502241\,m^{-1}$ and $k_2=503187\,m^{-1}$. (c) Manifests the plot of $P_{\nu_e\rightarrow\nu_{\mu}}$ against $t$ when $k_1=805122\,m^{-1}$ and $k_2=805770\,m^{-1}$. From the plots, we find that   oscillation frequency decreases with the increase in the energy of the mass eigenstates.}
\label{fig:2x2 relativistic Dirac}
\end{figure}

\begin{figure}[!]
\centering
    \subfigure[]{\includegraphics[width=0.24\textwidth]{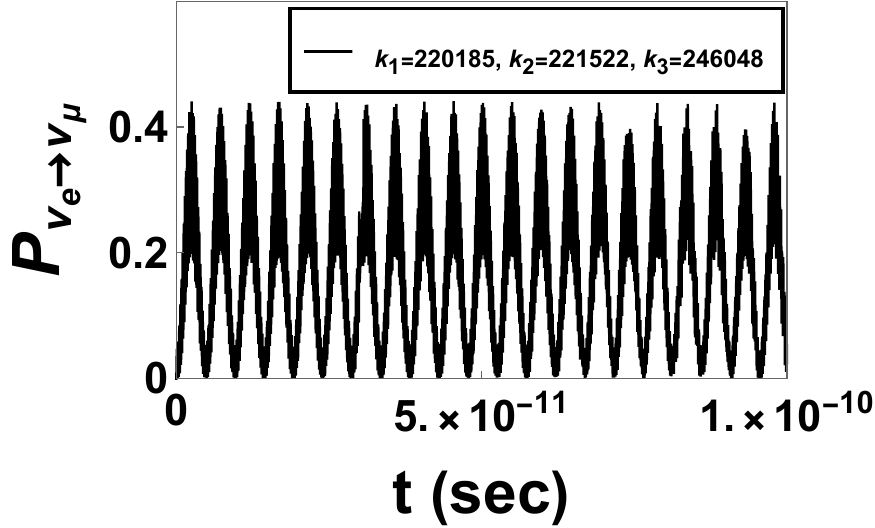}} 
    \subfigure[]{\includegraphics[width=0.24\textwidth]{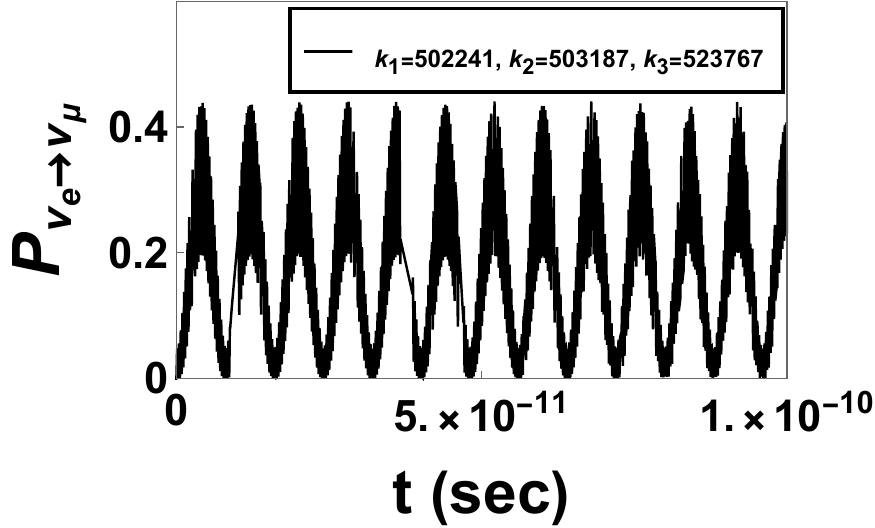}}
    \subfigure[]{\includegraphics[width=0.24\textwidth]{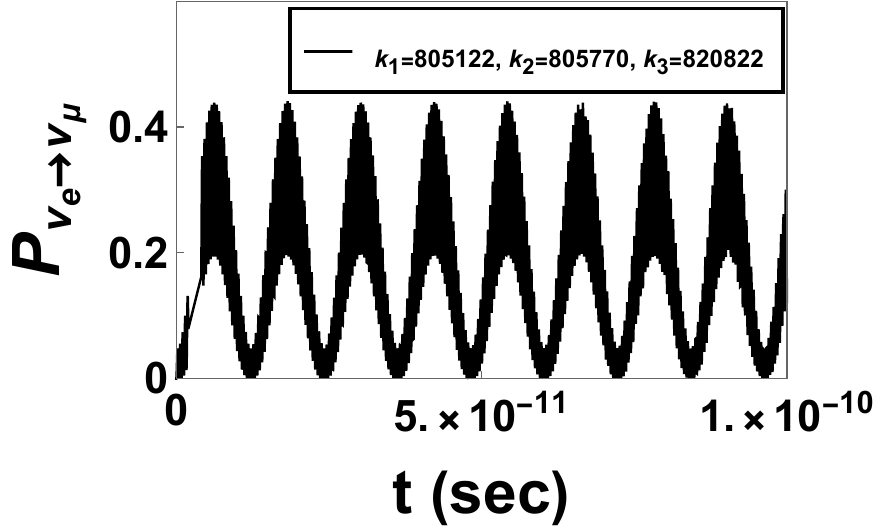}}  
 \caption{(a) The plot of $P_{\nu_e\rightarrow\nu_{\mu}}$ against $t$ is shown, where, $k_1=220185\,m^{-1}$, $k_2=221522\,m^{-1}$ and $k_3=246048\,m^{-1}$ for three flavor oscillation in relativistic scenario (Dirac). (b) Presents the plot of $P_{\nu_e\rightarrow\nu_{\mu}}$ against $t$ for $k_1=502241\,m^{-1}$, $k_2=503187\,m^{-1}$ and $k_3=523767\,m^{-1}$. (c) Manifests the plot of $P_{\nu_e\rightarrow\nu_{\mu}}$ against $t$ when $k_1=805122\,m^{-1}$, $k_2=805770\,m^{-1}$ and $n_3=820822\,m^{-1}$ respectively. From the plots, we perceive that the oscillation  frequency decreases when the  energy of the mass eigenstates increases.}
\label{fig:3x3 relativistic Dirac}
\end{figure}

Similar to Klein-Gordon scenario, it is observed that the oscillation frequency decreases as the mass eigenstates with higher $k$ values (see Figs.\,\ref{fig:2x2 relativistic Dirac}-\ref{fig:3x3 relativistic Dirac}). This cannot be explained with the help of the parameter $f'$ as two mass eigenstates have different masses. In this way, the two mass eigenstates cannot posses same wave number.

\begin{figure}[!]
\centering
    \subfigure[]{\includegraphics[width=0.24\textwidth]{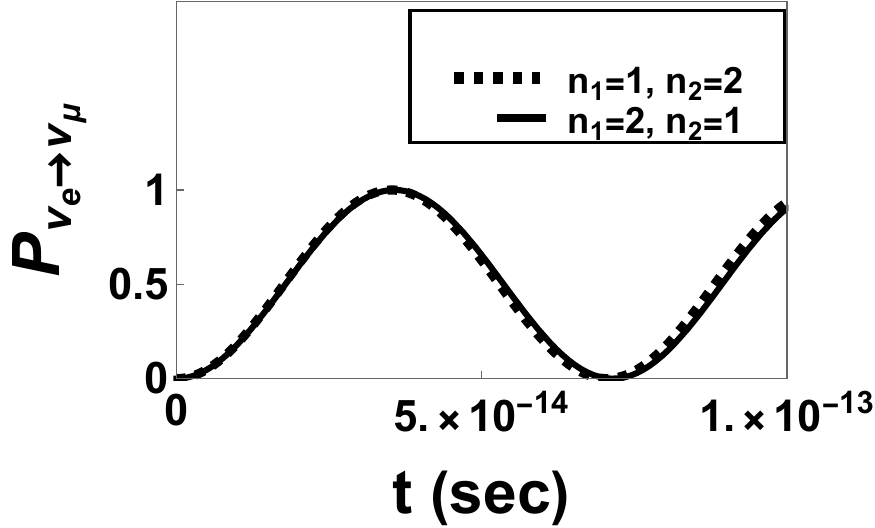}} 
    \subfigure[]{\includegraphics[width=0.24\textwidth]{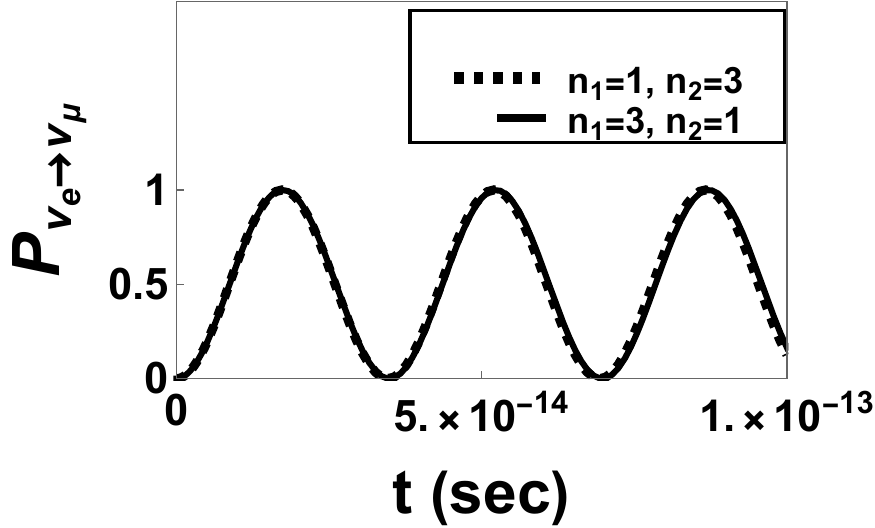}}
    \subfigure[]{\includegraphics[width=0.24\textwidth]{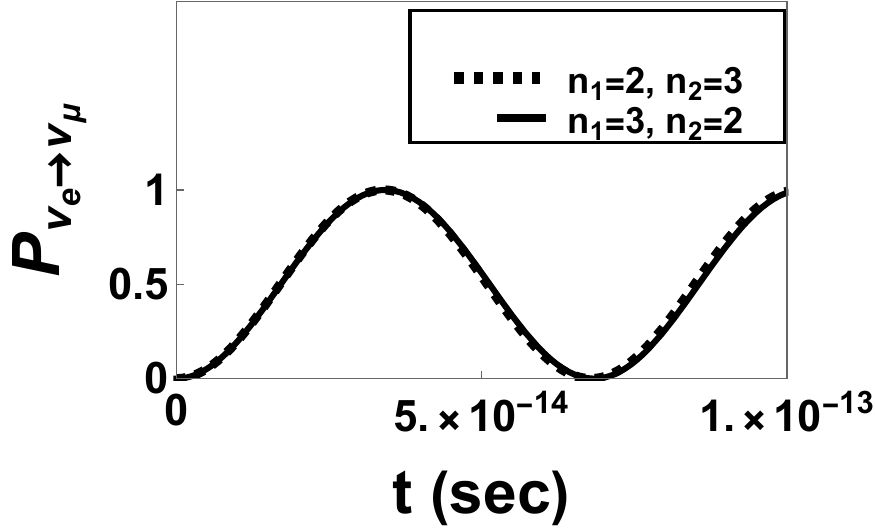}}
 \caption{(a) Gives the plot of $P_{\nu_e\rightarrow\nu_{\mu}}$ against $t$ for $n_1=1$ and $n_2=2$ and vice-versa when two flavor oscillation is considered in relativistic case (Klein-Gordon). (b) Presents the plot of $P_{\nu_e\rightarrow\nu_{\mu}}$ against $t$ for $n_1=1$ and $n_2=3$ and vice-versa. (c) Displays the plot of $P_{\nu_e\rightarrow\nu_{\mu}}$ against $t$, where, $n_1=2$ and $n_2=3$ and vice-versa. From the figure, we notice rarely any change in oscillation frequency for above mentioned energies.}
\label{fig:2x2 KG relativistic for differnt energy}
\end{figure}

\begin{figure}[!]
\centering
    \subfigure[]{\includegraphics[width=0.24\textwidth]{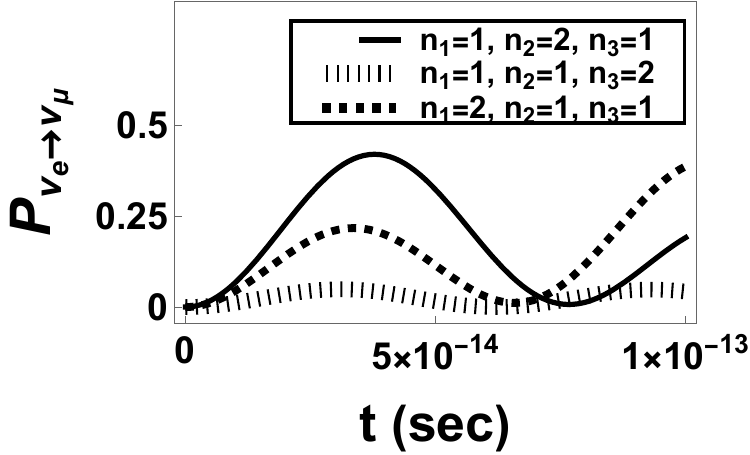}} 
    \subfigure[]{\includegraphics[width=0.24\textwidth]{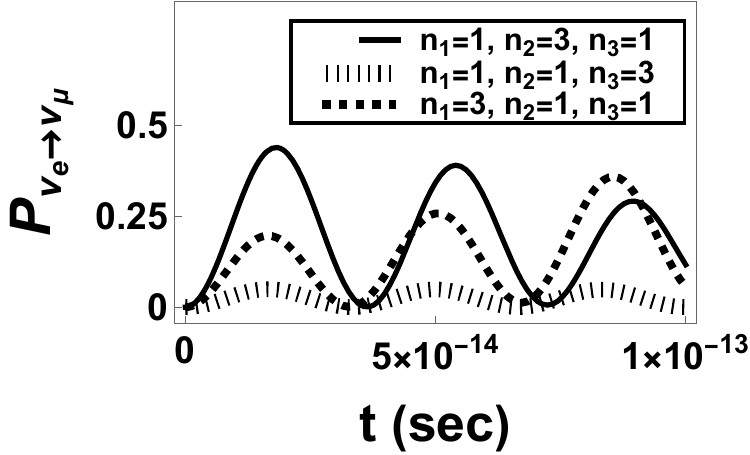}}
    \subfigure[]{\includegraphics[width=0.24\textwidth]{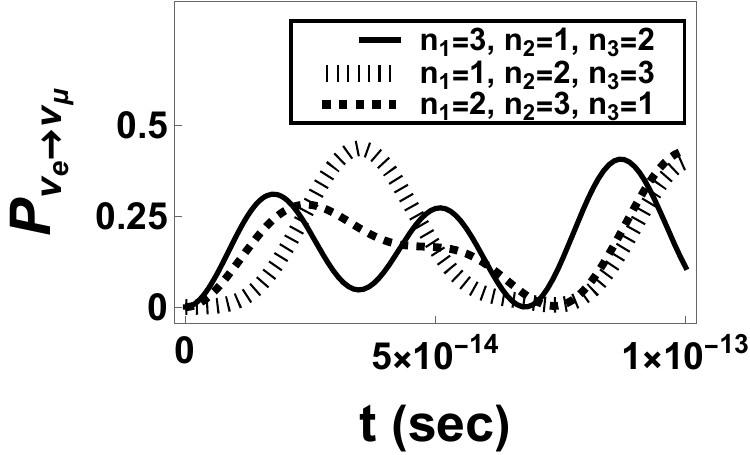}}
 \caption{(a) Presents the plot of $P_{\nu_e\rightarrow\nu_{\mu}}$ against $t$ when three flavor oscillation in relativistic case is considered (Klein-Gordon), where, $n_1$, $n_2$ and $n_3$ can have values $(1,1,2)$, $(1,2,1)$ and $(2,1,1)$ respectively. (b) Shows the plot of $P_{\nu_e\rightarrow\nu_{\mu}}$ against $t$ for the combinations of quantum numbers $(1,1,3)$, $(1,3,1)$ and $(3,1,1)$ respectively. (c) Plot of $P_{\nu_e\rightarrow\nu_{\mu}}$ against $t$ is manifested  where the values of $n_1$, $n_2$ and $n_3$ are chosen to be $(1,2,3)$, $(2,3,1)$ and $(3,1,2)$ respectively. From the plots, we observe that the frequency and the amplitude of oscillation changes with the change in energy of the mass eigenstates.}
\label{fig:3x3 KG non relativistic for differnt energy}
\end{figure}

\begin{figure}[!]
\centering
    \subfigure[]{\includegraphics[width=0.24\textwidth]{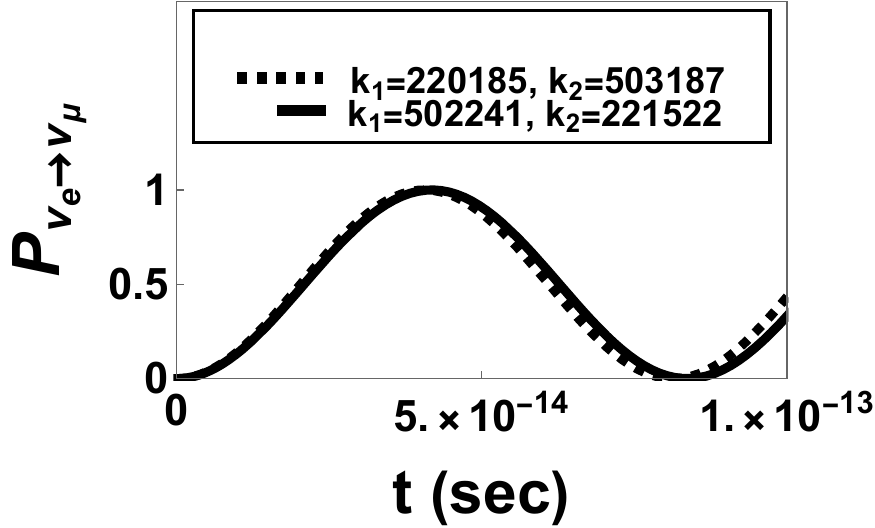}} 
    \subfigure[]{\includegraphics[width=0.24\textwidth]{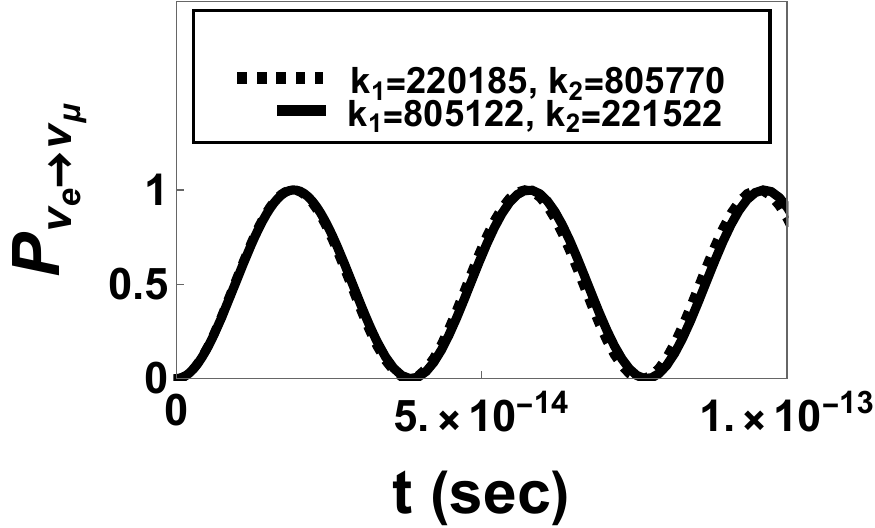}}
    \subfigure[]{\includegraphics[width=0.24\textwidth]{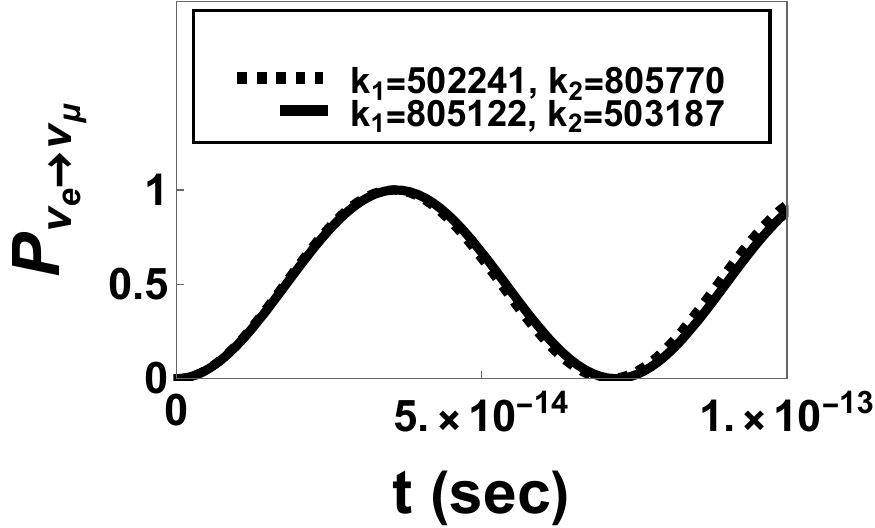}}
 \caption{(a) Manifests the plot of $P_{\nu_e\rightarrow\nu_{\mu}}$ against $t$ for the combinations  $k_1=220185\,m^{-1}$, $k_2=503187\,m^{-1}$ and $k_1=502241\,m^{-1}$, $k_2=221522\,m^{-1}$ when two flavor oscillation is considered in relativistic case (Dirac). (b) Displays the plot of $P_{\nu_e\rightarrow\nu_{\mu}}$ against $t$ for the combinations of the values of the wave number; $k_1=220185\,m^{-1}$, $k_2=805770\,m^{-1}$ and $k_1=805122\,m^{-1}$, $k_2=221522\,m^{-1}$. (c) Presents the plot of $P_{\nu_e\rightarrow\nu_{\mu}}$ against $t$, where, the wave number $k$ can have values $k_1=502241\,m^{-1}$, $k_2=805770\,m^{-1}$ and $k_1=805122\,m^{-1}$, $k_2=503187\,m^{-1}$. From the figure, we observe very small change in oscillation frequency for above mentioned energies.}
\label{fig:2x2 Dirac relativistic for differnt energy}
\end{figure}

\begin{figure}[!]
\centering
    \subfigure[]{\includegraphics[width=0.24\textwidth]{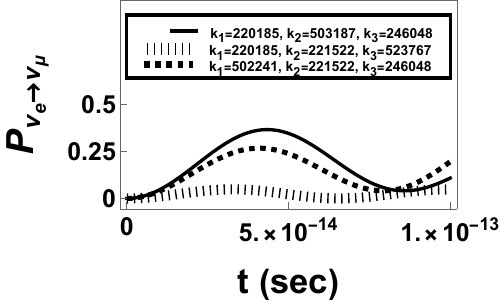}} 
    \subfigure[]{\includegraphics[width=0.24\textwidth]{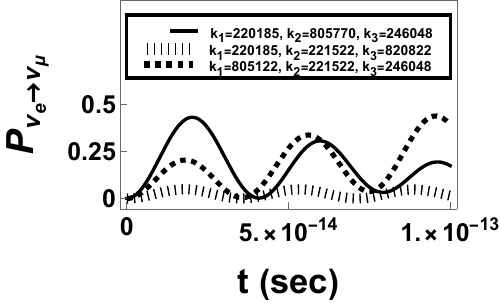}}
    \subfigure[]{\includegraphics[width=0.24\textwidth]{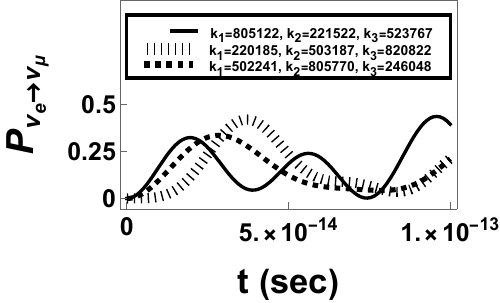}}
 \caption{(a) The plot of $P_{\nu_e\rightarrow\nu_{\mu}}$ against $t$ is shown for three flavor oscillation in relativistic case (Dirac) for combinations of the wave number ($k_1=220185\,m^{-1}$, $k_2=503187\,m^{-1}$, $k_3=246048\,m^{-1}$), ($k_1=220185\,m^{-1}$, $n_2=221522\,m^{-1}$, $k_3=523767\,m^{-1}$) and ($k_1=502241\,m^{-1}$, $k_2=221522\,m^{-1}$, $k_3=246048\,m^{-1}$). (b) Gives the plot of $P_{\nu_e\rightarrow\nu_{\mu}}$ against $t$ where the combinations of quantum numbers are ($k_1=220185\,m^{-1}$, $n_2=805770\,m^{-1}$, $k_3=246048\,m^{-1}$), ($k_1=220185\,m^{-1}$, $k_2=221522\,m^{-1}$, $k_3=820822\,m^{-1}$) and ($k_1=805122\,m^{-1}$, $k_2=221522\,m^{-1}$, $k_3=246048\,m^{-1}$). (c) Displays the plot of $P_{\nu_e\rightarrow\nu_{\mu}}$ against $t$, where, the values of $k_1$, $k_2$ and $k_3$ are chosen in the following combinations, ($k_1=805122\,m^{-1}$, $n_2=221522\,m^{-1}$, $k_3=523767\,m^{-1}$), ($k_1=220185\,m^{-1}$, $k_2=503187\,m^{-1}$, $k_3=820822\,m^{-1}$) and ($k_1=502241\,m^{-1}$, $k_2=805770\,m^{-1}$, $k_3=246048\,m^{-1}$). From the plots, we observe that the frequency and the amplitude of oscillation changes when the energy of the mass eigenstates changes.}
\label{fig:3x3 Dirac relativistic for differnt energy}
\end{figure}

For the \textbf{case-II}, we observe similar circumstances for Klein-Gordon and Dirac scenarios. It is seen that the amplitude and the oscillation frequency vary as the quantum numbers associated with the mass eigenstates are not same\,(see Figs.\,\ref{fig:2x2 KG relativistic for differnt energy}-\ref{fig:3x3 Dirac relativistic for differnt energy}). 

Further, if we look into the fact that neutrinos are electromagnetically neutral particles, they might resemble their own anti-particle state i.e., a Majorana Particle \cite{Arodz:2019kdq, Borsten:2016mfn}. However, the scenarios remain akin to the Dirac case as we perceive the same transcendental equation for the allowed wave number $k_i$ \cite{DeVincenzo:2019rmy}. Apart from that, we obtain plots of $P_{\nu_e\rightarrow\nu_{\mu}}$ against $t$ analogous to Dirac case.

 We highlight that we choose $\theta\,=\,45^\circ$ for two neutrino case while for three neutrino case, we choose the best fit values of the mixing angles as per the oscillation data (see Table \ref{table:Values Of Parameters in NO}) \cite{Esteban:2020cvm, Gonzalez-Garcia:2021dve}.

\subsection{Conditions for No Flavour Oscillation}

In the theory of standard neutrino oscillation, where the neutrino travels in free space, the oscillation stops when $\Delta m^2_{21}=0$. This signifies that the neutrino mass eigenvalues cannot be degenerate. If the two mass eigenstates $\nu_1$ and $\nu_2$ have got same masses then its impossible to distinguish between the two states and under this scenario, there arises no question of mixing and hence the neutrino oscillation disappears. However, we see that if neutrino is confined within a well, then $\nu_1$ and $\nu_2$ can be discriminated by virtue of the quantum numbers $n_1$ and $n_2$, in addition to their distinct masses $m_1$ and $m_2$ respectively. So, for the sake of discussion, even if we may speculate a situation when $m_1=m_2$, the two states will still remain distinguishable unless $n_1=n_2$ and thus, mixing as well as flavour oscillation do occur. Unlike the confined case, for free space, the said speculation will neither lead to mixing nor to oscillation.  However, for a confined neutrino, the oscillation will not occur under some  specific conditions which are discussed below. 

\begin{figure}[!]
\centering
    \subfigure[]{\includegraphics[width=0.24\textwidth]{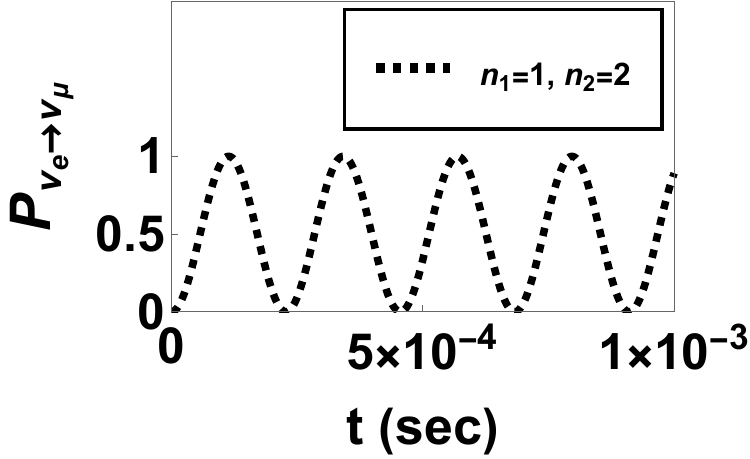}\label{fig:degenerate 2x2 Schrodinger}} 
    \subfigure[]{\includegraphics[width=0.24\textwidth]{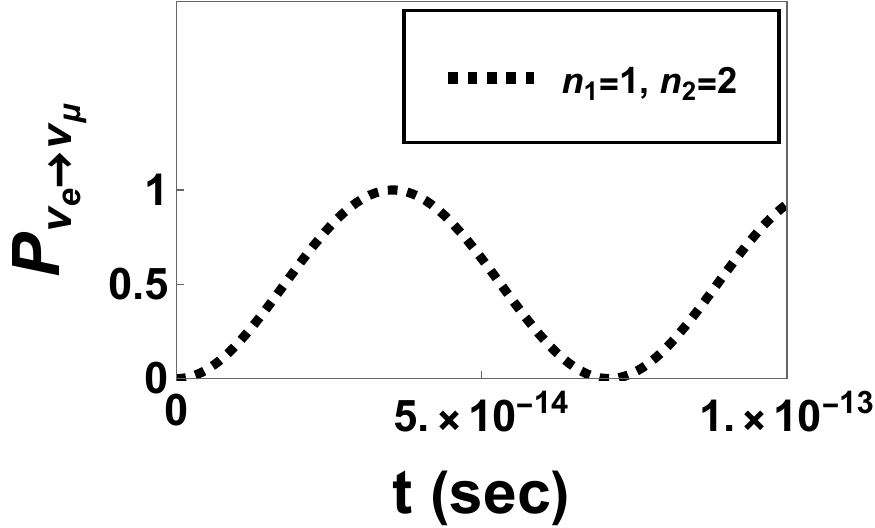}\label{fig:degenerate 2x2 KG}}
    \subfigure[]{\includegraphics[width=0.24\textwidth]{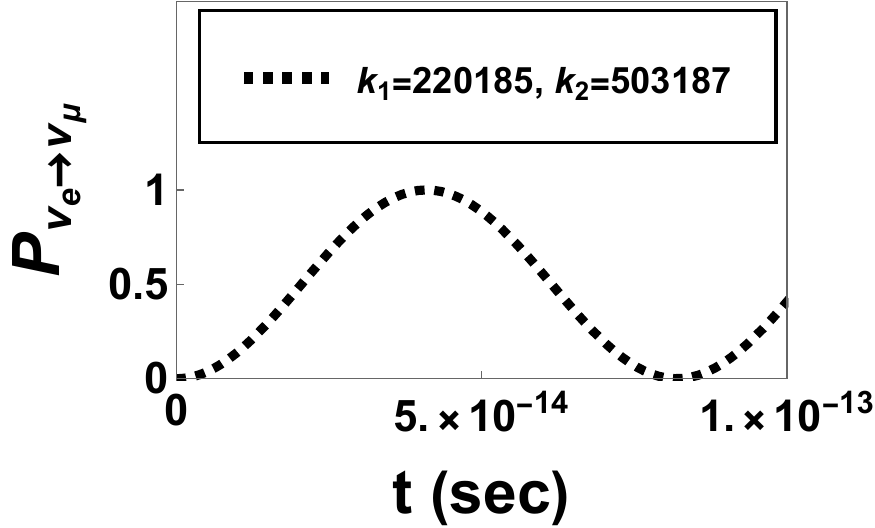}\label{fig:degenerate 2x2 Dirac}}
 \caption{(a) Presents the plot of $P_{\nu_e\rightarrow\nu_{\mu}}$ against $t$ for $m_1=m_2$ in case of two flavor oscillation in non-relativistic case. (b) Displays the plot of $P_{\nu_e\rightarrow\nu_{\mu}}$ against $t$ for two flavor oscillation in relativistic case (Klein-Gordon), where, $m_1=m_2$. (c) Shows the plot of $P_{\nu_e\rightarrow\nu_{\mu}}$ against $t$ for $m_1=m_2$ for two flavor oscillation in relativistic case (Dirac). From the plots, we notice a non-zero oscillation probability even if the mass eigenvalues are degenerate.}
\label{fig:degenerate 2x2}
\end{figure}

\begin{figure}[!]
\centering
    \subfigure[]{\includegraphics[width=0.24\textwidth]{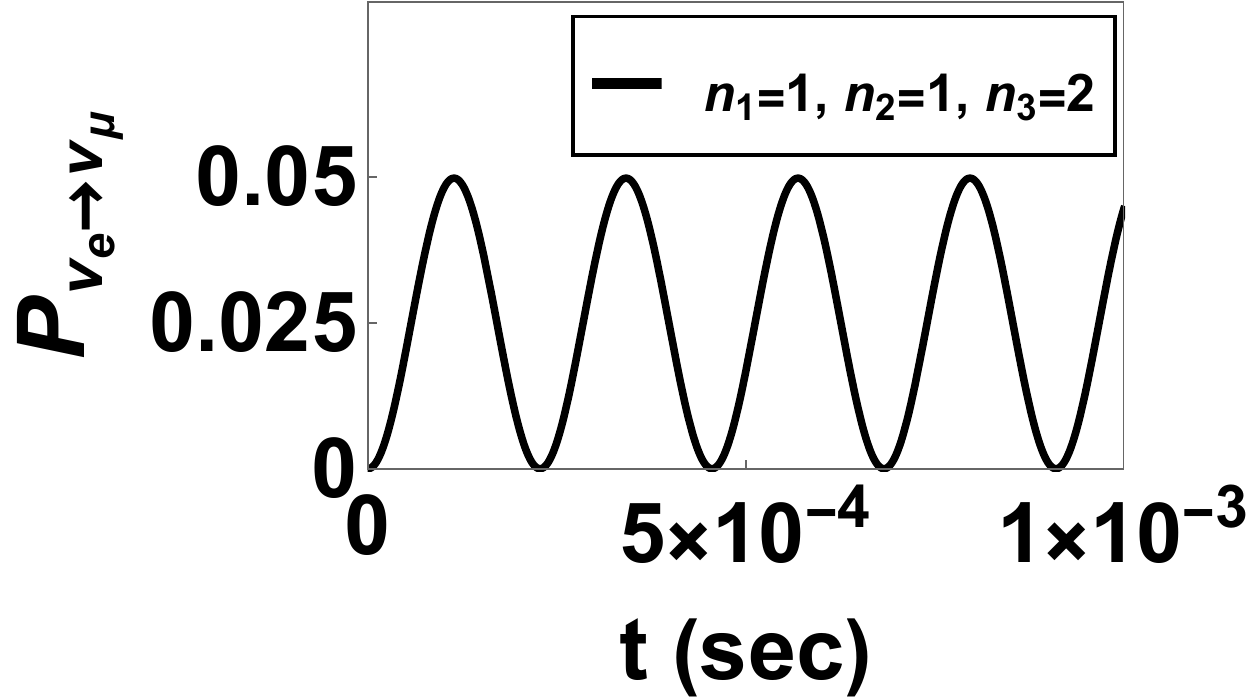}\label{fig:degenerate 3x3 Schrodinger}} 
    \subfigure[]{\includegraphics[width=0.24\textwidth]{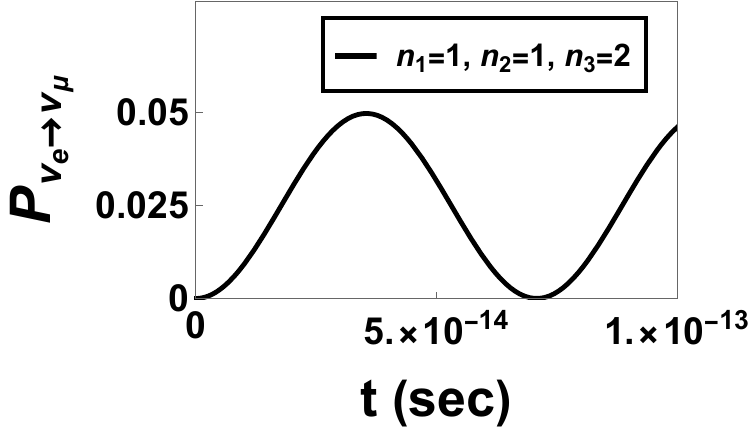}\label{fig:degenerate 3x3 KG}}
    \subfigure[]{\includegraphics[width=0.24\textwidth]{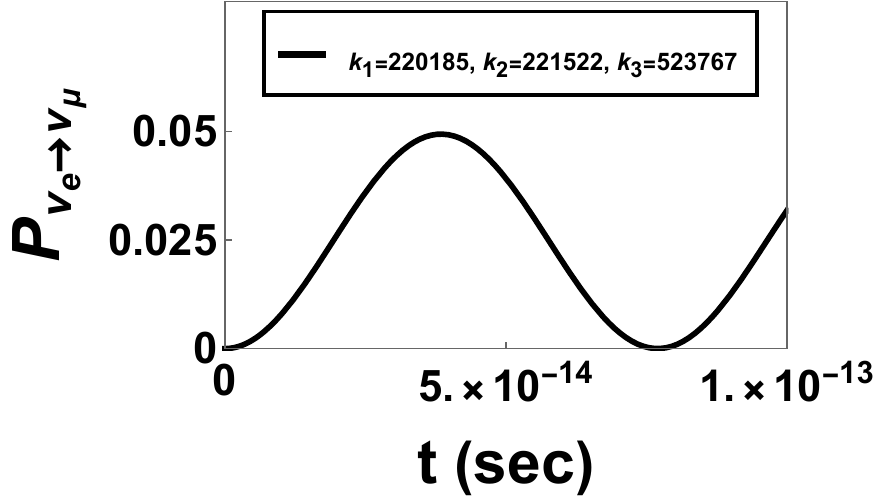}\label{fig:degenerate 3x3 Dirac}}
 \caption{(a) The plot of $P_{\nu_e\rightarrow\nu_{\mu}}$ against $t$ is manifested for $m_1=m_2=m_3$, when three flavor oscillation in non-relativistic case is considered. (b) Presents the plot of $P_{\nu_e\rightarrow\nu_{\mu}}$ against $t$ where, $m_1=m_2=m_3$ for three flavor oscillation in relativistic case (Klein-Gordon). (c) Displays the plot of $P_{\nu_e\rightarrow\nu_{\mu}}$ against $t$ for three flavor oscillation in relativistic case (Dirac) for $m_1=m_2=m_3$. From the plots, we see that the non-zero oscillation probability exists for degenerate neutrino mass eigenvalues.}
\label{fig:degenerate 3x3}
\end{figure}

In non-relativistic scenario, we exploit Eq.\,(\ref{2x2 probability}) and Eq.\,(\ref{3x3 probability}) to draw a important consequence. If two neutrino mass eigenstates posses different quantum number ($n_1 \neq n_2$), we witness non zero value of $P_{\nu_e\rightarrow\nu_\mu}$ even if the neutrino mass eigenstates are degenerate i.e., $m_1=m_2$ (see Figs. \ref{fig:degenerate 2x2 Schrodinger} and \ref{fig:degenerate 3x3 Schrodinger}). 

From Eq.\,(\ref{2x2 probability}), it is seen that the oscillation will not occur (two neutrino case) for the following condition,

\begin{equation}
m_1\,=\,\Bigg(\frac{n^2_1}{n^2_2}\Bigg)\,m_2.
\label{2x2 condition}
\end{equation}

For three neutrino framework, the condition for zero oscillation probability is shown below,

\begin{equation}
m_3\,=\,\Bigg(\frac{n^2_3}{n^2_1}\Bigg)\,m_1\,=\,\Bigg(\frac{n^2_3}{n^2_2}\Bigg)\,m_2.
\end{equation}

For the relativistic scenario, the obtained results are similar results to the non-relativistic case (see Figs. \ref{fig:degenerate 2x2 KG}-\ref{fig:degenerate 2x2 Dirac} and \ref{fig:degenerate 3x3 KG}-\ref{fig:degenerate 3x3 Dirac}). However, the oscillation will not occur for some specific conditions.

For the Klein-Gordon scenario, there will be no oscillation if,

\begin{equation}
\Delta\,m^2_{21}\,=\,\frac{\pi^2\hslash^2}{L^2c^2}\,(n^2_1-n^2_2),
\label{2x2 KG condition}
\end{equation}

in view of the two neutrino framework, whereas for three neutrino case, the neutrino oscillation vanishes if,

\begin{equation}
\frac{\Delta m^2_{21}}{(n^2_1-n^2_2)}=\frac{\Delta m^2_{31}}{(n^2_1-n^2_3)}=\frac{\Delta m^2_{32}}{(n^2_2-n^2_3)}= \Bigg(\frac{\pi^2 \hslash^2}{L^2 c^2}\Bigg).
\end{equation}

Similarly, for the Dirac case, we exploit Eq.\,(\ref{2x2 probability for Dirac case}) and perceive the condition for no oscillation as shown below,

\begin{equation}
\Delta\,m^2_{21}\,=\,\frac{\hslash^2}{c^2}\,(k^2_1-k^2_2),
\label{2x2 Dirac condition}
\end{equation}

for two neutrino case and the same for three neutrino framework, is obtained as in the following,

\begin{equation}
\frac{\Delta m^2_{21}}{(k^2_1-k^2_2)}=\frac{\Delta m^2_{31}}{(k^2_1-k^2_3)}=\frac{\Delta m^2_{32}}{(k^2_2-k^2_3)}=\Bigg(\frac{\hslash^2}{c^2}\Bigg).
\end{equation}

It is observed that the quantum numbers associated with the mass eigenstates play a crucial role in the neutrino flavour conversion when the neutrino is trapped inside a infinite square well potential.

\section{Summary and Discussion \label{section 5}}

In our work, we follow the Pontecorvo's approach and set a thought experiment to study the neutrino oscillation inside a confined region. We formulate our work by considering both relativistic and non-relativistic frameworks. The oscillation is expressed with respect to time and a few important consequences are discussed. We have studied the probability of flavour conversion $P_{\nu_e\rightarrow\nu_\mu}$ considering two cases where the mass eigenstates $\nu_1$ and $\nu_2$ are attributed to the same quantum number and the case when the quantum numbers are different. For the first case, we encounter a significant difference between the oscillation frequency for relativistic and non-relativistic scenarios. For the relativistic scenario, it is seen that the oscillation frequency decreases as the mass eigenstates corresponds to excited energy states while for the non-relativistic scenario, the results are completely opposite. For the second case, we observe a change in the oscillation frequency when the energies of the mass eigenstates change. However, for the three neutrino case, a change in both frequency and amplitude of oscillation is observed when the mass eigenstates are placed in higher energy states. Apart from this, we discuss some conditions for which the neutrino oscillation will not occur for both the relativistic and non-relativistic cases. A similar attempt is made by applying the wave packet approach in the non-relativistic framework \cite{Johns:2019amh}. But in the present work, we deal with different equations of motions and explore the relativistic scenario as well.

It is needless to mention that the present work is based on a speculation. The neutrinos have got very low interaction cross-section with matter and the world is almost transparent to them. So, confining the neutrinos inside a potential well is not at all straight-forward. However, there may exist some scenarios where a very dense object like the core of a neutron star may act as an obstacle to neutrinos and thus, the latter may get trapped there for several seconds \cite{Reddy:1996tw, Stuchlik:2011zzb}. In this dense environment, the neutrino-neutrino interactions may trigger the self induced flavour conversion of neutrinos \cite{Mirizzi:2015eza}. So, in this context, we hope that our study may find some relevance.

\section{Acknowledgement} 

The research work of Pralay Chakraborty is supported by Innovation in Science Pursuit for Inspired Research (INSPIRE), Department of Science and Technology, Government of India, New Delhi vide grant No. IF190651. 

\biboptions{sort&compress}

\appendix

\section{Important Derivations \label{appendix 1}}

To derive the probability expression Eq.\,(\ref{2x2 probability}), we would like to start from the Pontecorvo's approach of finding the oscillation probability.

Inside the well, the wave functions at $t=0$,

\begin{eqnarray}
\mid\nu_1(t=0)\rangle\,&=&\,\sqrt{\frac{2}{L}}\,\sin\Big(\frac{n_1 \pi x}{L}\Big), \\
\mid\nu_2(t=0)\rangle\,&=&\,\sqrt{\frac{2}{L}}\,\sin\Big(\frac{n_2 \pi x}{L}\Big),
\end{eqnarray}

For two neutrino scenario, the neutrino mixing is depicted as shown in the following

\begin{equation}
 \begin{bmatrix}
 \mid\nu_{e}\rangle\\
 \mid\nu_{\mu}\rangle\\
 \end{bmatrix}
 =
 \begin{bmatrix}
 \cos\theta & \sin\theta\\
 -\,\sin\theta & \cos\theta\\
 \end{bmatrix}
 \begin{bmatrix}
 \mid\nu_1\rangle\\
 \mid\nu_2\rangle\\
 \end{bmatrix},
 \label{2x2 mixing}
 \end{equation}
where, $\theta$ is the mixing angle.

At $t=0$, we consider the neutrino is purely electron type. In this regard, the wave function corresponds to $\nu_e$ is given by,

\begin{equation}
\mid\psi(t=0)\rangle=\cos\theta\mid\nu_1(t=0)\rangle+\sin\theta\mid\nu_2(t=0)\rangle.
\end{equation}

The mass eigenstates $\mid\nu_1(t=0)\rangle$ and  $\mid\nu_2(t=0)\rangle$ will evolve in time $t$ as shown below: 
\begin{eqnarray}
\mid\nu_k(t)\rangle= e^{-\frac{i}{\hslash}\,\hat{H_k} t}\mid\nu_k(t=0)\rangle = e^{-\frac{i}{\hslash}\,E_{n_k} t}\mid\nu_k(t=0)\rangle,\quad\quad k=1,2.
\end{eqnarray}

At time $t$,

\begin{equation}
\mid\psi(t)\rangle=\cos\theta e^{-\frac{i}{\hslash} E_{n_1}t}\mid\nu_1(t=0)\rangle +\sin\theta e^{-\frac{i}{\hslash} E_{n_2}t}\mid\nu_2(t=0)\rangle.
\label{state at t}
\end{equation}

From Eq.\,(\ref{2x2 mixing}), we can write,

\begin{eqnarray}
\mid\nu_1(t=0)\rangle &=& \cos\theta \mid \nu_e\rangle -\sin\theta \mid \nu_{\mu}\rangle,\\
\mid\nu_2(t=0)\rangle &=& \sin\theta \mid \nu_e\rangle +\cos\theta \mid \nu_{\mu}\rangle,
\end{eqnarray}

and we rewrite Eq.\,(\ref{state at t}) as shown in the following,

\begin{eqnarray}
\mid\psi(t)\rangle &=& \cos\theta(\cos\theta \mid\nu_e\rangle-\sin\theta \mid\nu_{\mu}\rangle) e^{-\frac{i}{\hslash} E_{n_1}t} +\sin\theta(\sin\theta \mid\nu_e\rangle+\cos\theta \mid\nu_{\mu}\rangle) e^{-\frac{i}{\hslash} E_{n_2}t},\nonumber\\
&=& e^{-\frac{i}{\hslash} E_{n_1}t} (\cos^2 \theta+ \sin^2 \theta e^{\frac{i}{\hslash} (E_{n_1}-E_{n_2})t}) \mid\nu_e\rangle-e^{-\frac{i}{\hslash} E_{n_1}t} (1-e^{\frac{i}{\hslash} (E_{n_1}-E_{n_2})t})\cos\theta\sin\theta\mid\nu_{\mu}\rangle.\nonumber\\
&=& c_e\mid\nu_e\rangle+c_\mu\mid\nu_\mu\rangle.
\label{complete wave function}
\end{eqnarray}

where, $c_e=\langle \nu_e\mid\psi(t)\rangle$ and $c_\mu=\langle \nu_\mu\mid\psi(t)\rangle$.

Now, the probability that the electron neutrino $\nu_e$ oscillate into muon neutrino $\nu_\mu$ is given by,

\begin{eqnarray}
P(\nu_e-\nu_\mu)&=&c_\mu c^*_\mu \nonumber\\
&=& (1-e^{\frac{i}{\hslash} (E_{n_1}-E_{n_2})t})(1-e^{-\frac{i}{\hslash} (E_{n_1}-E_{n_2})t})\cos^2\theta\sin^2\theta\nonumber\\
&=&\frac{1}{4}(2-2\cos\frac{(E_{n_1}-E_{n_2})t}{\hslash})\sin^2 2\theta\nonumber\\
&=&\sin^2 2\theta\, \sin^2 \frac{(E_{n_1}-E_{n_2})t}{2\hslash}
\label{general probability}
\end{eqnarray}

From Eq.\,(\ref{quantised energy}), we put the expressions of $E_{n_1}$ and $E_{n_2}$ in Eq.\,(\ref{general probability}) and obtain the expression for $P(\nu_e-\nu_\mu)$ as shown in the following,

\begin{equation}
P_{\nu_e\rightarrow\nu_{\mu}}\,=\,\sin^2(2\,\theta)\,\sin^2 \Big[\frac{\pi^2\,\hslash\,t}{4\,L^2}\Big(\frac{n_1^2}{m_1}-\frac{n_2^2}{m_2}\Big)\Big].
\end{equation}

We would like to highlight an important discussion here. In principle, the mass eigenstate $\nu_{1}$ may correspond to any of the states like $\nu_1^1$, $\nu_{1}^2$, $\nu_1^3\cdots$ for $n_1=1,2,3,\cdots$ respectively and the same holds good for $\nu_2$ as well. Under this scenario, the mixing of neutrinos can be visualised as shown below.

\begin{eqnarray}
\nu_e &=& c_1^1 \nu_1^1 + c_1^2 \nu_1^2 + c_1^3 \nu_1^3+ \cdots+ c_2^1 \nu_2^1 + c_2^2 \nu_2^2 + c_2^3 \nu_2^3+\cdots\\
\nu_\mu &=& d_1^1 \nu_1^1 + d_1^2 \nu_1^2 + d_1^3 \nu_1^3+ \cdots+ d_2^1 \nu_2^1 + d_2^2 \nu_2^2 + d_2^3 \nu_2^3+\cdots
\end{eqnarray} 

Hence,
\begin{equation}
\begin{pmatrix}
\nu_e\\
\nu_\mu
\end{pmatrix} = \tilde{U} \begin{pmatrix}
\nu^{1}_1\\ \nu^{2}_1\\ \nu^{3}_1\\ \vdots \\ \nu^{1}_2 \\ \nu^{2}_2 \\ \nu^{3}_2\\ \vdots
\end{pmatrix}.
\end{equation}

We understand that $\tilde{U}$ is a rectangular matrix. If we believe that the neutrino mixing matrix is unitary, then certainly at a time, only two states can contribute towards the mixing. Now this mixing can happen in three ways: \textbf{(i)} mixing between two states having same mass but different quantum numbers ($m_1=m_2,\,n_1\neq n_2$) \textbf{(ii)} mixing between two states having same quantum number but different masses\,($n_1=n_2,\,m_1\neq m_2$) and \textbf{(iii)} that between the states having different quantum numbers and different masses\,($m_1\neq m_2,\,n_1\neq n_2$).

\pagebreak

\begin{table}[!]
\centering
\begin{tabular}{ccc} 
\hline
Parameters & btf\,$\pm$\,$1\,\sigma$ & $3\,\sigma$\,range\\ 
\hline\hline
$\theta_{12}/^{\circ}$ & $33.44^{+0.77}_{-0.74}$ & 31.27\,-\,35.86\\
\hline
$\theta_{23}/^{\circ}$ & $49.2^{+1.0}_{-1.3}$ & 39.5\,-\,52\\
\hline
$\theta_{13}/^{\circ}$ & $8.57^{+0.13}_{-0.12}$ & 8.20\,-\,8.97\\
\hline
$\delta_{CP}/^{\circ}$ & $194^{+52}_{-25}$ & 105\,-\,405\\
\hline
$\frac{\Delta\,m_{21}^2}{10^{-5}\,eV^2}$ & $7.42^{+0.21}_{-0.20}$ & 6.82\,-\,8.04\\
\hline
$\frac{|\Delta\,m_{31}^2|}{10^{-3}\,eV^2}$ & $2.515^{+0.028}_{-0.027}$ & 2.431\,-\,2.599\\
\hline
\end{tabular}
\caption{$3\,\nu$ oscillation parameters obtained from  global analysis of neutrino data in normal ordering \cite{Esteban:2020cvm}.} 
\label{table:Values Of Parameters in NO}
\end{table}

\begin{table}
\centering
\begin{tabular}{cccc} 
\hline
Mass Eigenstates & $k_1 (\,m^{-1})$ & $k_2(\,m^{-1})$ & $k_3(\,m^{-1})$\\ 
\hline\hline
$m_1$ & $220185$ & $502241$ & $805122$\\
\hline
$m_2$ & $221522$ & $503187$ & $805770$\\
\hline
$m_3$ & $246048$ & $523767$ & $820822$\\
\hline
\end{tabular}
\caption{Numerical values of $k_1$, $k_2$ and $k_3$ with respect to $m_1$, $m_2$ and $m_3$ respectively.} 
\label{table:Numerical values}
\end{table}

\end{document}